\documentclass[10pt,letterpaper]{article}
\usepackage[top=0.85in,footskip=0.75in]{geometry}

\usepackage{amsmath,amssymb}

\usepackage{changepage}

\usepackage[utf8x]{inputenc}

\usepackage{textcomp,marvosym}

\usepackage{cite}

\usepackage{nameref,hyperref}

\usepackage[right]{lineno}

\usepackage{microtype}
\DisableLigatures[f]{encoding = *, family = * }

\usepackage[table]{xcolor}

\usepackage{array}

\newcolumntype{+}{!{\vrule width 2pt}}

\newlength\savedwidth

\usepackage{setspace} 

\usepackage[aboveskip=1pt,labelfont=bf,labelsep=period,justification=raggedright,singlelinecheck=off]{caption}

\makeatletter
\renewcommand{\@biblabel}[1]{\quad#1.}
\makeatother

\date{}

\usepackage{lastpage,fancyhdr,graphicx}
\usepackage{epstopdf}

\usepackage{booktabs}
\usepackage{pdfpages}
\usepackage{pdflscape}                 
\usepackage{ulem}	
\usepackage{cooltooltips}
\usepackage{tikz}
\newcommand*\circled[1]{\tikz[baseline=(char.base)]{
		\node[shape=circle,draw,inner sep=2pt] (char) {#1};}}
\newcommand*{\inactivates}[0]{\tikz[baseline=-0.65ex]{ \draw [-|] (0,0) -- (2ex,0) ;}}

\begin{document}
\vspace*{0.2in}

\begin{flushleft}
{\Large
\textbf\newline
{Feedbacks from the metabolic network to the genetic network reveal regulatory modules in \textit{E. coli} and \textit{B. subtilis}}
}
\newline
\\
Santhust Kumar\textsuperscript{1},
Saurabh Mahajan\textsuperscript{2},
Sanjay Jain\textsuperscript{1,3,*}
\\
\bigskip
\textbf{1} Department of Physics and Astrophysics, University of Delhi, Delhi 110007, India
\\
\textbf{2} National Centre for Biological Sciences, Bangalore, Karnataka 560065, India
\\
\textbf{3} Santa Fe Institute, 1399 Hyde Park Road, Santa Fe, NM 87501, USA
\\
\bigskip
%
%





* E-mail: jain@physics.du.ac.in
\end{flushleft}


\section*{Abstract}
The genetic regulatory network (GRN) plays a key role in controlling the response of the cell to changes in the environment. Although the structure of GRNs has been the subject of many studies, their large scale structure in the light of feedbacks from the metabolic network (MN) has received relatively little attention. Here we study the causal structure of the GRNs, namely the chain of influence of one component on the other, taking into account feedback from the MN. 
First we consider the GRNs of \textit{E. coli} and \textit{B. subtilis} without feedback from MN and illustrate their causal structure. Next we augment the GRNs with feedback from their respective MNs by including (a) 
links from genes coding for enzymes to metabolites produced or consumed in reactions catalyzed by those enzymes and (b) links from metabolites to genes coding for transcription factors whose transcriptional activity the metabolites alter by binding to them.
We find that the inclusion of feedback from MN into GRN significantly affects its causal structure, in particular the number of levels and relative positions of nodes in the hierarchy, and the number and size of the strongly connected components (SCCs). 
We then study the functional significance of the SCCs. For this we identify condition specific feedbacks from the MN into the GRN by retaining only those enzymes that are essential for growth in specific environmental conditions simulated via the technique of flux balance analysis (FBA). We find that the SCCs of the GRN augmented by these feedbacks can be ascribed specific functional roles in the organism. Our algorithmic approach thus reveals relatively autonomous subsystems with specific functionality, or regulatory modules in the organism. This automated approach could be useful in identifying biologically relevant modules in other organisms for which network data is available, but whose biology is less well studied.


\section{Introduction}
The control of gene expression is central to cellular dynamics. All cellular processes -- metabolism, growth, cell division, response to stimuli, etc. -- are related to characteristic subsets of genes that need to be expressed for the respective processes to take place inside the cell. The expression of genes inside a cell is controlled via a complex process of inter-regulation involving, in part, a network of genes called the gene regulatory network (GRN). Due to the large number of genes and interactions in a GRN and the presence of feedback loops, it becomes difficult to track the causal regulation from one component of the GRN to another. In order to obtain a system level understanding of gene interactions of an organism, an understanding of the structure of its GRN and its design principles is necessary.

Several studies have characterized the GRNs on a global and local scale. The GRNs have been found to follow an exponential in-degree distribution and a power law out-degree distribution \cite{thieffry_specific_1998,guelzim_topological_2002,lee_transcriptional_2002}. They have been shown to possess a hierarchical and modular organization \cite{ma_hierarchical_2004,yu_genomic_2006,farkas_topological_2006,cosentino_lagomarsino_hierarchy_2007,samal_regulatory_2008,freyre-gonzalez_functional_2008,jothi_genomic_2009,rodriguez-caso_basic_2009,bhardwaj_analysis_2010,corominas-murtra_origins_2013}. While hierarchy depicts the regulatory flow of information in a system, identifying the modules in a system has also been found useful in comprehending its functional and structural organization \cite{hartwell_molecular_1999,oltvai_lifes_2002,segal_modular_2003,wagner_road_2007}. Modules have been defined in various ways: as a group of genes that express together \cite{eisen_cluster_1998,segal_module_2003,ihmels_defining_2004,ihmels_revealing_2002}, as a cluster or community of nodes that connect tightly together compared to other nodes in the network \cite{girvan_community_2002,newman_finding_2004}, as a set of connected nodes revealed upon the removal of common global regulators \cite{ma_decomposition_2004}, and as network motifs \cite{alon_network_2007,shen-orr_network_2002} which have a specific functionality. 

Despite its frequent isolated treatment the GRN of an organism functions in conjunction with other networks in the cell, in particular the metabolic network (MN). In order to have a better understanding of the organisational structure and functioning of the GRN, consideration of the influence of the MN over it is crucial. Several works study the GRN and MN in an integrated manner to explore various structural, dynamical or evolutionary aspects of their interplay \cite{ihmels_principles_2004,covert_integrating_2004,herrgard_integrated_2006,yeang_joint_2006,shlomi_genome-scale_2007,samal_regulatory_2008,seshasayee_principles_2009}. 
Recently a combined network of GRN, MN and protein interactions has been used to explore the cascading impact of perturbations originating in different parts of the combined network \cite{klosik_interdependent_2017}. However, the global architecture of the integrated GRN and MN remains to be elucidated. Here we study some aspects of this architecture, in particular focusing on the effect of feedback from MN to GRN on the hierarchical and modular structure involved in the overall control of metabolism.


To this end, we obtain the data about the genetic regulatory interactions and metabolism of two microorganisms, \textit{E. coli} and \textit{B. subtilis}, from a number of sources in  literature \cite{salgado_regulondb_2013,freyre-gonzalez_lessons_2013,sierro_dbtbs:_2008,reed_expanded_2003,henry_ibsu1103:_2009,keseler_ecocyc:_2013,goelzer_reconstruction_2008}. Certain metabolites bind to transcription factors (TFs) and alter their gene regulatory activity. We add to the GRN nodes corresponding to such metabolites as well as links from these nodes to the genes coding for TFs to which the metabolites bind. In addition we include links to these metabolites from genes coding for enzymes which catalyze the reactions of these metabolites. The augmented GRNs so obtained consist of approximately 3300 nodes and 9300 edges  (\textit{E. coli}), and 1700 nodes and 3500 edges (\textit{B. subtilis}).

We first organize each GRN (without including the metabolite nodes) into a hierarchical structure by using a modified form of the vertex sort algorithm of Jothi et al \cite{jothi_genomic_2009} which involves finding strongly connected components (SCCs) of the network, thereby elucidating its regions of feedback and causal structure.

Second, we augment the GRN by introducing nodes and links corresponding to metabolites belonging to the MN as described above, and discuss how the hierarchical structure changes. We show that GRNs retain their characteristic hierarchical structure upon including the feedbacks from metabolism; however they become more complicated and the causal ordering governed by the relative positions of genes among the various levels in the hierarchy is significantly altered.

Third, we exploit the fact that not all feedbacks from the MN are functional at a given time. To this end we simplify the augmented network by first identifying functionally relevant feedbacks from the MN into the GRN under a number of simulated environmental conditions (ECs) using the technique of flux balance analysis (FBA) \cite{orth_what_2010}, and then augmenting the GRN with only these functionally relevant feedbacks. The structure of the augmented network thus obtained is easier to interpret in terms of dynamics and biological function. We then interpret the resulting SCCs of the network as modules, following Rodriguez-Caso et al \cite{rodriguez-caso_basic_2009}. Our definition of modules differs somewhat from that of \cite{rodriguez-caso_basic_2009} in that we do not limit the module to just the SCC, but also consider the proximal regulatory circuit around the nodes belonging to the SCC. This in conjunction with the fact that metabolic nodes are also part of our network allows us to assign a specific functional role to most of the SCCs. This role follows from the circuit diagram of the SCC and the manner in which it is embedded in the full network, which makes its qualitative dynamics and biological function quite evident. This effectively provides a list of biologically relevant dynamical sub-systems of the cell for further investigation. In addition to finding and classifying the important modules of the joint GRN and MN of \textit{E. coli} and \textit{B. subtilis}, our methodology provides an algorithmic approach for finding important sub-systems of organisms for which the GRN and MN has been obtained.

Finally we also attempt to find sub-modules of the largest SCC of {\it E. coli} which is large and complicated and cannot be assigned a simple functional role.

\section{The GRN of \textit{E. coli} and \textit{B. subtilis}}\label{sec:chap3_full_grn}
We obtain the gene regulatory network interaction data of \textit{E. coli} from RegulonDB \cite{salgado_regulondb_2013}, and that of \textit{B. subtilis} from Freyre-Gonzalez et al \cite{freyre-gonzalez_lessons_2013} which is a curated database of regulatory interactions based on DBTBS, a database of transcriptional regulation in \textit{B. subtilis} \cite{sierro_dbtbs:_2008}. 
The GRN of \textit{E. coli} contains 3277 nodes and 8740 edges. 
The GRN of \textit{B. subtilis} 
consists of 1681 nodes and 3096 edges. 
Further details are given in Table \ref{table:grn_and_mn_numbers}.
For reference purposes, we designate these GRNs (without consideration of feedback from metabolism) as graph $ \mathcal{G_A} $. The GRNs, without any knowledge of hierarchy, are pictured in Fig \ref{fig:graphA_all}A. The entire data of graph $ \mathcal{G_A} $ (list of various types of nodes, links, hierarchical level, etc.) is given in Supporting Information (SI) file S1.

\begin{table}[t!]	
	\caption[Overview \textit{B. subtilis} and \textit{E. coli} networks.]{\textbf{Networks overview.} Overview of the various \textit{E. coli} and \textit{B. subtilis} networks used in the study.}
	\label{table:grn_and_mn_numbers}
	\begin{tabular}{p{0.35\linewidth}p{0.35\linewidth}rr}
		\toprule
		&  & \textit{E. coli} & \textit{B. subtilis} \\ 
		\midrule
		Gene Regulatory Network ($ \mathcal{G_A} $) &  &  &  \\ 
		& Total Genes & 3277 & 1681 \\ 
		& Total regulator genes & 248 & 154 \\ 
		& $\sigma$-factor genes & 7 & 14 \\ 
		& Transcription factor genes & 191 & 126 \\ 
		& Regulating ncRNA genes & 50 & 14 \\
		& Regulated-only genes* & 3029 & 1527 \\ 
		& Interactions & 8740 & 3096 \\ 
		& Levels in Hierarchy & 7 & 7 \\ 
		& Number of SCCs & 7 & 4 \\ 
		& Size of largest SCC & 56 & 6 \\ 
		Metabolic Network &  &  &  \\ 
		& Genes & 904 & 1103 \\ 
		& Metabolites & 761 & 1139 \\ 
		& Reactions & 931 & 1437 \\ 
		GRN with allosteric feedbacks from Metabolic network ($ \mathcal{G_B} $)
		&  &  &  \\ 
		& Nodes & 3343 & 1710 \\ 
		& Gene Nodes & 3277 & 1681 \\ 
		& Metabolite Nodes & 66 & 29 \\ 
		& Total edges & 9279 & 3546 \\ 
		& Gene to gene edges & 8740 & 3096\\
		& Gene to metabolite edges & 462 & 416\\
		& Metabolite to gene edges & 77 & 34\\
		& Levels in Hierarchy & 7 & 11 \\ 
		& Number of SCCs & 20 & 9 \\ 
		& Size largest SCC & 378 & 85 \\  
		\bottomrule
	\end{tabular}
	\begin{flushleft}
		* Genes whose product does not directly regulate any other gene (such nodes have no outgoing links in the GRN, but receive incoming links from regulator nodes.)
	\end{flushleft}
\end{table}

\subsection{The hierarchical structure}
The GRNs of bacteria are known to have a pyramidal hierarchical structure where the number of nodes in the hierarchical levels decreases as one moves to the higher levels \cite{ma_decomposition_2004,yu_genomic_2006,cosentino_lagomarsino_hierarchy_2007,samal_regulatory_2008,jothi_genomic_2009,bhardwaj_analysis_2010,kumar_analysis_2015}. However, with the growth in the size of the networks in recent years through the addition of new regulatory interactions, feedbacks have emerged. Feedbacks disturb the simple chain of command architecture and are critical because they give rise to properties like homeostasis, bistability etc.
Our effort is directed towards the understanding of global organization and causal relationships between different regulatory elements of the GRN. Therefore we first remove the self loops so that we can focus on inter-nodal feedbacks that exist in the network. We capture the feedbacks in the network obtained after removing self loops by identifying its strongly connected components (SCCs). A SCC of a directed graph is a maximal subgraph, such that for any pair of nodes $ i $ and $ j $ in the subgraph, there exists a path in the sub-graph from $ i $ to $ j $ and from $ j $ to $ i $.
Together the SCCs capture all the feedbacks in the network. Furthermore, a SCC, owing to its property that any node belonging to it can affect any of its other nodes, can be thought of as forming a semi-autonomous group of nodes, or a module, of the full network. Understanding the functioning of these modules and placing them within the parent network can provide a way to understand the functional architecture of the parent network \cite{rodriguez-caso_basic_2009}.

\begin{figure}[h!]
	\vspace*{-1cm}
	\begin{center}
	\includegraphics[width=0.70\textwidth]{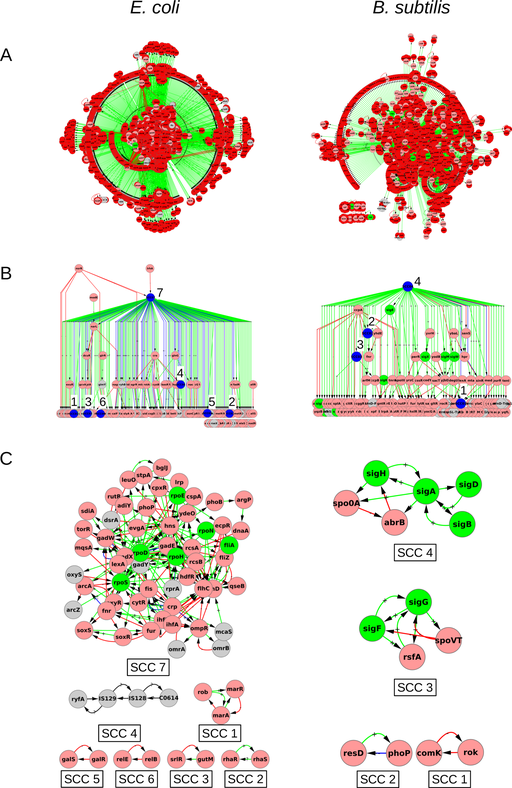}
	\end{center}
	\caption[GRNs of \textit{E. coli} and \textit{B. subtilis}: graph $ \mathcal{G_A} $]{\textbf{GRNs of \textit{E. coli} and \textit{B. subtilis}: graph $ \mathcal{G_A} $.} \textbf{A.} The whole GRNs of \textit{E. coli} and \textit{B. subtilis} pictured using one common layout. \textbf{B.} The revealed hierarchical structure (condensed graph) of the GRNs. Only the genetic regulators are shown. The regulated-only genes (nodes with no outgoing links) have not been shown for purposes of clarity; they would have been placed below the lowest level shown of the hierarchy, constituting a level-0. In other words, only level 1 and higher of the hierarchy are shown. Node colors---Blue: strongly connected components (SCCs), Pink: transcription factor genes (TFs), Red: genes coding for enzymes, Green: $ \sigma $-factor genes, grey: non-coding RNAs (ncRNAs). Edge colors represent the sign of regulation---Green ($ + $): activating, Red ($ - $): inhibiting, Blue ($ +- $ or $ -+ $): dual, Black (?): interaction mode remains uncharacterised. The sign ($ + $, $ - $, $ +- $, ?) is also marked on the link. The SCCs have been numbered in the hierarchy, with their composition shown in C. \textbf{C.} The SCCs of the GRNs of \textit{E. coli} and \textit{B. subtilis}. Networks are pictured using Cytoscape \cite{shannon_cytoscape:_2003}. Higher resolution images of this and other figures in the paper that allow node names to be read can be made available upon request from the authors. Images used here are downsized to meet the file size requirements.}
	\label{fig:graphA_all}
\end{figure}

Varying procedures of defining the hierarchical levels for a GRN exist in the literature \cite{ma_decomposition_2004,yu_genomic_2006,jothi_genomic_2009,bhardwaj_analysis_2010}. Here, we use a combination of the procedure proposed by Jothi et al \cite{jothi_genomic_2009} along with an intuitive bottom-up placement of hierarchical levels proposed by Yu et al \cite{yu_genomic_2006}, to construct the hierarchical levels (see Methods section \ref{methods_hierarchical_organisation} for details).
Briefly, we perform a condensation of the GRNs of \textit{E. coli} and \textit{B. subtilis}, which produces corresponding directed acyclic graphs (DAGs). The condensation of a graph involves identification of SCCs of the graph, replacing each of the SCC by a single node (SCC node), and replacing the edges to/from the original nodes in the SCCs by new edges to/from the SCC node. The DAG, by construction, is devoid of cycles or feedback loops, and hence it is possible to unambiguously place the nodes of the network in a chain of command hierarchy. The hierarchical organization of the GRN of \textit{E. coli} and \textit{B. subtilis} is shown in Fig \ref{fig:graphA_all}B.

The hierarchical organization of the GRNs obtained through the procedure above clearly elucidates the causal regulatory relationships between the transcriptional regulators as well as the regions of feedbacks (SCCs) in the GRNs. In the Fig. \ref{fig:graphA_all}B, all the regulatory links point downwards as a node in a higher level can regulate a node in a lower level but no node in the lower level can regulate a node in the higher level. This depicts the flow of information or regulatory influence in a GRN. We see that the hierarchy of transcriptional regulators in the GRNs have only a few localized islands of feedback.


\subsection{The inter-genic feedbacks}
The GRN of \textit{E. coli} has a total of 7 SCCs while the  GRN of \textit{B. subtilis} has a total of 4 SCCs (Fig. \ref{fig:graphA_all}C). Most SCCs in both organisms are small, having between 2 and 4 nodes. The largest SCC (LSCC) in \textit{E. coli} (SCC 7 in Fig. \ref{fig:graphA_all}C) has 56 nodes, and is much larger than the second largest SCC which has only 4 nodes. In \textit{B. subtilis}, however, the largest SCC has only 6 nodes and is not much larger than the second largest SCC. It is likely that this feature is a consequence of the fact that the present database of \textit{B. subtilis} has far less coverage of the actual network compared to \textit{E. coli}. It may be noted that only 1681  genes in \textit{B. subtilis} are represented in the present network compared to 3277 in \textit{E. coli} (see table \ref{table:grn_and_mn_numbers}) while the total number of genes in both organisms is comparable and in the range 4000-4500. The average degree of the present GRN for \textit{B. subtilis} (1.84) is also smaller than for \textit{E. coli} (2.67) suggesting that many links may not have been documented. An earlier version of the \textit{E. coli} network also did not have a giant SCC (the largest SCC in \cite{rodriguez-caso_basic_2009} had 11 nodes). Thus it is possible that with greater coverage of the \textit{B. subtilis} network in the future a larger giant SCC may emerge in the \textit{B. subtilis} network.

Despite these differences in size, the largest SCCs of both the organisms have two striking similarities. First, they are located at the top of the hierarchy. Second, both consist of global regulatory factors. The largest SCC in \textit{E. coli} contains 6 (of 7) $ \sigma $-factors and global TFs like \textit{crp}, \textit{fis}, \textit{ihf}, \textit{hns}. The largest SCC in \textit{B. subtilis} contains the housekeeping sigma factor \textit{sigA}. Quantitatively, the largest SCC of \textit{E. coli} has over $ 80\% $ of the top $ 10 $ global regulators (based upon out-degree of the genes), while \textit{B. subtilis} has over $ 25\% $.
It is clear from the respective set of SCCs that, for the present constructions of GRNs, the GRN of \textit{E. coli} has more cycles than that of \textit{B. subtilis} (on account of more SCCs and larger size of the LSCC). 
The above suggests that the GRNs of both bacteria have largely hierarchical tree like structures with mainly small and isolated feedback loops, and one larger feedback structure located near the top of the hierarchy that influences a large number of downstream genes. This picture is essentially the same as in \cite{rodriguez-caso_basic_2009} except that we find a much larger LSCC in \textit{E. coli} at the top of the hierarchy. The LSCC in \textit{E. coli} in the present study is 14 times larger than the second largest SCC. Some of the smaller SCCs in \cite{rodriguez-caso_basic_2009} now have coalesced into one and form part of the present LSCC.

\section{Feedbacks from metabolic network into GRNs}\label{sec:chap3_feedbacks_from_MN_into_GRN}
\begin{figure}[h!] 
	\centering
	\vspace*{-1cm}
	\includegraphics[width=.7\textwidth]{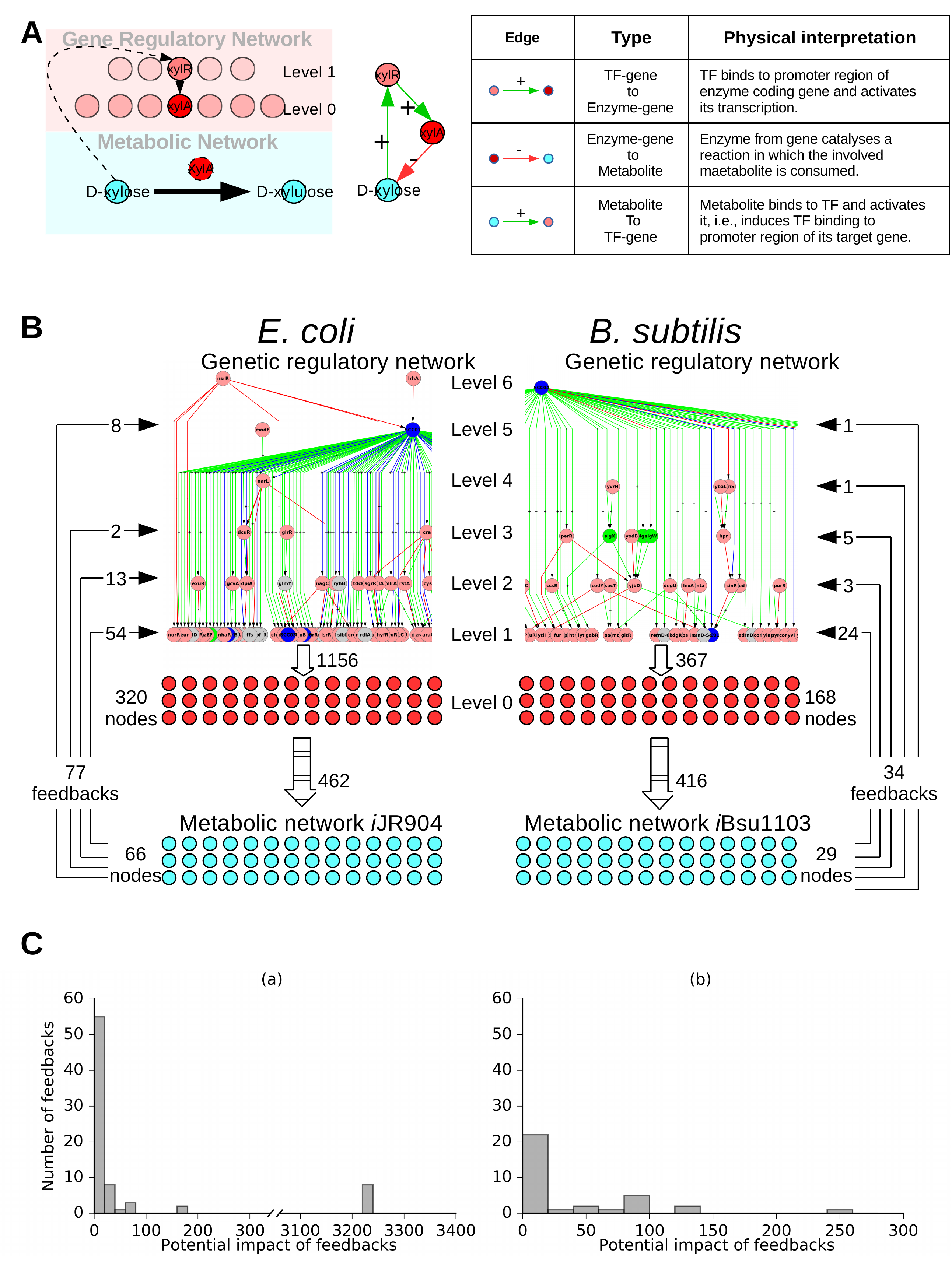}
	\caption[Schematic of feedbacks into GRN from MN]{\textbf{Schematic of feedbacks into GRN from MN.} \textbf{A.} Example of the feedback from the MN to the GRN. The TF \textit{XylR} (coded for by the gene \textit{xylR}), when activated, binds to and switches on the gene \textit{xylA} (which codes for the enzyme \textit{XylA}). \textit{XylA} catalyzes the metabolic reaction in which metabolite D-xylose is converted into D-xylulose. D-xylose binds to \textit{XylR} and activates it namely, causes it to bind to its target gene). The two figures in A are representation of the above statements. The table categorizes each type of link involved and gives its physical interpretation. This feedback has the dynamical consequence that excess of D-xylose in the cell is regulated by enhancing its conversion into D-xylulose (through activation of \textit{XylR} that up-regulates \textit{XylA} which catalyzes the conversion). \textbf{B.} GRNs with feedbacks from metabolic network. \textbf{C.}  Histograms (bin size 20) of potential impact of feedbacks from metabolic network into the respective GRNs of \textit{E. coli}, (a), and \textit{B. subtilis}, (b). Note that to avoid clutter we do not introduce separate nodes for a TF and the gene that codes for it. Thus the node representing a TF coding gene does double duty and represents both the genes and the corresponding TF. An arrow from a metabolite node (cyan) to a TF coding gene (pink) does not mean that the metabolite increases the expression of the gene. It means that the metabolite activates the TF coded for by the gene (namely, causes the TF to bind to its target genes and carry out its function---activating or inhibiting transcription of its target gene). However a green/red arrow ($ + $ or $ - $) from a TF coding gene (pink) to another TF coding gene (as in GRN of panel B) means that the former TF activates/inactivates the transcription of the latter gene.}
	\label{fig:schematic_link_interpretation_and_grn_feedback_from_mn}
\end{figure}

The activity of some TFs is controlled by the binding of small molecules that are part of the cell's metabolism. These metabolites typically bind to the TF to alter its capacity to bind to target promoters. The production/consumption of such metabolites may be catalysed by enzymes whose genes are regulated by the same TF. 
Each binding reaction of a metabolite with a TF that alters the latter's regulatory activity thus  creates a feedback from the MN to the GRN (see example in Fig. \ref{fig:schematic_link_interpretation_and_grn_feedback_from_mn}A).
We gather the information pertaining to the enzymatic catalysis of reaction from the metabolic models of the bacteria \cite{reed_expanded_2003,henry_ibsu1103:_2009}, the TF metabolite interaction from RegulonDB \cite{salgado_regulondb_2001,salgado_regulondb_2013}, Ecocyc \cite{keseler_ecocyc:_2013} and Goelzer et al \cite{goelzer_reconstruction_2008}, and integrate this information with the GRN. For the purposes of reference, we label the GRN augmented with feedbacks from respective metabolic networks as graph $ \mathcal{G_B} $ (details in Methods section \ref{methods_grn_with_feedback_from_metabolism_gb}; schematic in Fig. \ref{fig:schematic_link_interpretation_and_grn_feedback_from_mn}B).

Feedbacks from the metabolic network can be employed to effect a range of changes in the GRN: from small to large. The impact a feedback has depends upon the level of the GRN into which it feeds. The higher the level a feedback goes into, the larger the potential impact. Fig. \ref{fig:schematic_link_interpretation_and_grn_feedback_from_mn}B shows that most of the feedbacks from metabolic network into GRN are at lowest level.
We explored the potential impact of feedbacks from metabolic network into GRN. One can define the potential impact of a feedback from a metabolite to a TF as just the number of genes downstream of that TF (at any distance).
A histogram of the potential impact of feedbacks is shown in Fig. \ref{fig:schematic_link_interpretation_and_grn_feedback_from_mn}C. 
As expected from Fig. \ref{fig:schematic_link_interpretation_and_grn_feedback_from_mn}B, Fig. \ref{fig:schematic_link_interpretation_and_grn_feedback_from_mn}C shows that most of the allosteric feedbacks from metabolic networks into GRNs are employed to affect only a small set of genes, and a few feedbacks are also employed to bring about global changes. Interestingly, the feedbacks from metabolic networks into GRNs do not seem to have intermediate levels of potential impact (see Fig. \ref{fig:schematic_link_interpretation_and_grn_feedback_from_mn}C).
It would be interesting to await more data for \textit{B. subtilis}, wherein the LSCC and perhaps feedback to it could emerge.

Feedbacks with large impact can possibly be related to effecting global changes, while those with low impact are likely to be effecting fine-tuning changes. 
An example of a feedback having a global impact is from the metabolite cAMP (cyclic adenosine mono phosphate) which binds the TF \textit{Crp} (coded by gene \textit{crp}) in the core of the \textit{E. coli} network (it is one of the eight mentioned in Fig. \ref{fig:schematic_link_interpretation_and_grn_feedback_from_mn}B).
It is well known that the absence of readily metabolizable carbon sources, such as glucose, results in the increased level of cAMP, which through its binding to \textit{Crp} influences a large number of genes to change the state of the cell \cite{robison_comprehensive_1998,zheng_identification_2004,grainger_studies_2005}. On the other hand the D-xylose feedback, described in Fig. \ref{fig:schematic_link_interpretation_and_grn_feedback_from_mn}A, is at level 1. Its effect is more local. Instead of making a global impact, the feedback only affects a few genes like \textit{xylA} involved in D-xylose metabolism with an effect to restore D-xylose balance. (A list of metabolites, the TFs they bind to and the level of the TF are given in SI file \nameref{S2_graphB}.)
Level wise distribution of the feedbacks, Fig. \ref{fig:schematic_link_interpretation_and_grn_feedback_from_mn}B and Fig. \ref{fig:schematic_link_interpretation_and_grn_feedback_from_mn}C, indicates that a majority of the allosteric feedbacks are utilized for fine-tuning purposes.

\section{Structure and organization of GRN with metabolic feedbacks}\label{sec:chap3_graph_gB}
The incorporation of feedbacks into the GRN from the metabolic network brings substantial changes in the structure and organization of the GRN. For the purpose of reference, we label the GRN augmented with feedbacks from metabolism as graph $ \mathcal{G_B} $.
We probe the structure and organization of graph $ \mathcal{G_B} $ in the same algorithmic way as done previously for $ \mathcal{G_A} $, i.e., via strongly connected components and hierarchical structure of the corresponding $ \mathcal{G_B} $ condensed graphs. We first describe the SCC, and then the hierarchical structure. The graph $ \mathcal{G_B} $ is depicted in Figs. \ref{fig:graphB_sccs_ec_bs} and \ref{fig:graphB_hierarchical_structure_ec_bs}.

\subsection{Strongly connected components} 
\begin{figure}[h!] 
	\centering
	\includegraphics[width=.65\textwidth,trim={0 0 0 0},clip]{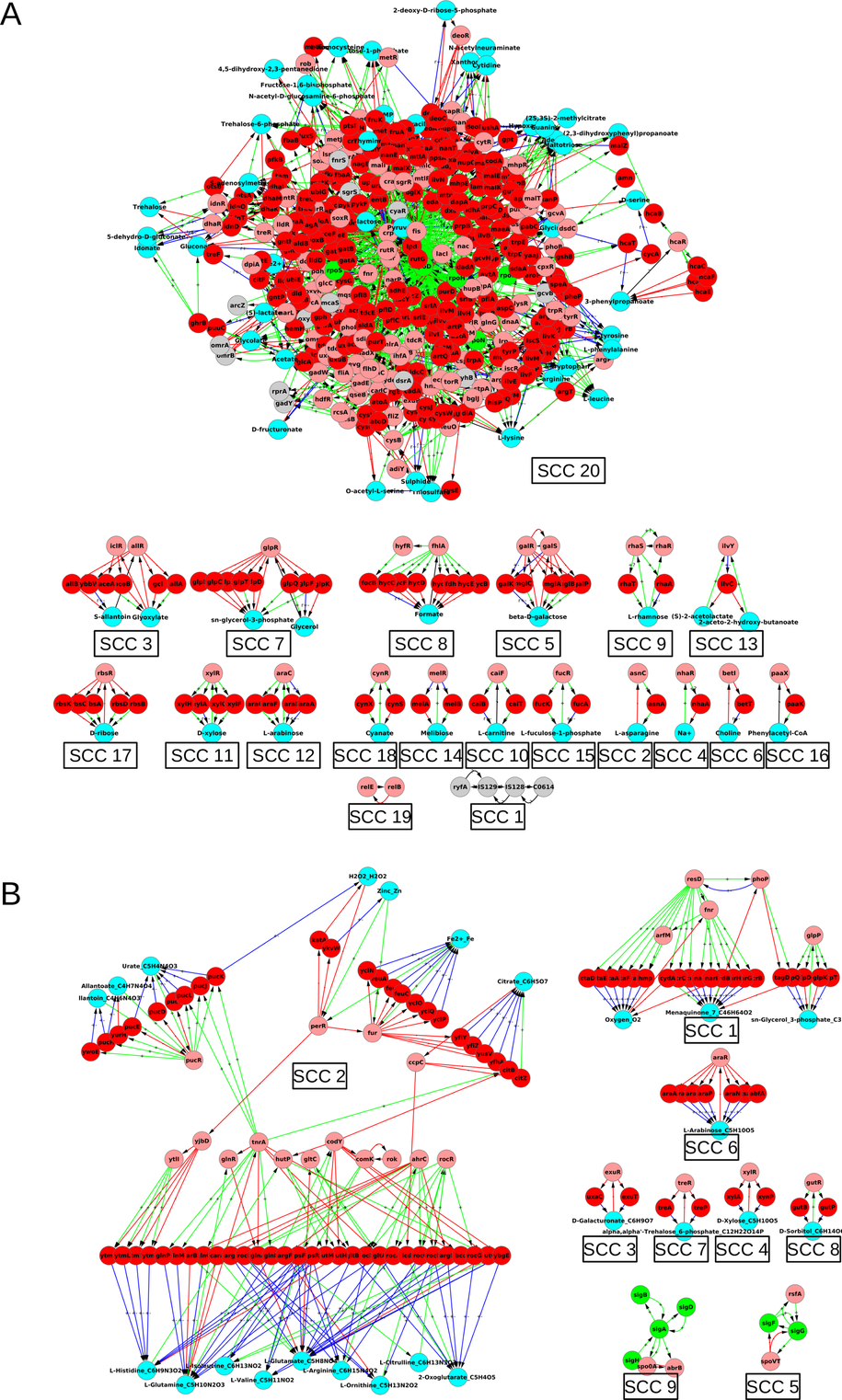}
	\caption[Strongly connected components of GRNs (graph $ \mathcal{G_B} $) of \textit{E. coli} and \textit{B. subtilis} with feedbacks from respective metabolic networks: graph $ \mathcal{G_B} $.]{\textbf{Strongly connected components of GRNs of \textit{E. coli} (A) and \textit{B. subtilis} (B) with feedbacks from respective metabolic networks: graph $ \mathcal{G_B} $.}  Colour code: Nodes---Red: genes coding for enzymes, Cyan: metabolites, rest same as in Fig. \ref{fig:graphA_all}; Edges---same as in Figs. \ref{fig:graphA_all} and \ref{fig:schematic_link_interpretation_and_grn_feedback_from_mn}. In addition to the convention described in Fig. \ref{fig:schematic_link_interpretation_and_grn_feedback_from_mn}A, we mention that a red arrow from a metabolite node to a TF coding gene node means that the metabolite binds to the TF and inactivates it, i.e., prevents the TF from binding to its target. Further a green arrow from an enzyme coding gene to a metabolite means that the enzyme catalyzes a reaction in which the metabolite is produced. In addition to the convention described in Fig. \ref{fig:schematic_link_interpretation_and_grn_feedback_from_mn}A, we mention that a red arrow from a metabolite node to a TF coding gene node means that the metabolite binds to the TF and inactivates it, i.e., prevents the TF from binding to its target. Further a green arrow from an enzyme coding gene to a metabolite means that the enzyme catalyzes a reaction in which the metabolite is produced.}
	\label{fig:graphB_sccs_ec_bs}
\end{figure}

Due to addition of feedbacks involving enzymes and metabolites, the largest SCCs were 6 fold and 14 fold bigger than SCCs of pure GRN (\textit{E. coli} and \textit{B. subtilis} respectively).
For \textit{E. coli}, increase in number of feedbacks further complicates the already complex structure of the largest SCC. 
In both the organisms, the amino acids are predominantly present in the largest SCCs, whereas the sugars mostly occupy the small SCCs. This makes sense with the following logic: The sub-networks involving the sugars function at the input end of the metabolic network. When a sugar is present as food in the environment, the corresponding sub-network is active, while sub-networks related to other sugars are inactive. Thus the sub-networks relating to the sugar molecules form individual SCCs. Once the food is taken in, a core machinery which is active in every simulated environmental condition can produce various required molecules, e.g., the building block amino acids, for the cell. We will later (section \ref{sec:chap4_identifying_function_modules}) describe results which will shed more light on this observation.
One stark difference between the largest SCCs of the two organisms relates to the presence of $ \sigma $-factors in the LSCC of \textit{E. coli} while the $ \sigma $-factors are absent in the LSCC of \textit{B. subtilis}. On the other hand there is also a similarity: amino acid and peroxide/iron modules are in the largest SCCs for both.
For \textit{B. subtilis} the new largest SCC can be loosely organized into 3 communities, Fig. \ref{fig:graphB_sccs_ec_bs}B, related to (a) one enriched in amino acids, (b) allantoin metabolism, and (c) with three sub-structures related to peroxide response, iron uptake and citrate metabolism. However, the LSCC (378 nodes) of \textit{E. coli} is densely connected and its decomposition into communities is not that obvious. We discuss some aspects of this later (section \ref{sec:chap4_structure_of_core}).
Further, the number of small SCCs has grown substantially both in size and in numbers (\textit{E. coli}: 6 to 19, and \textit{B. subtilis}: 4 to 8).

\subsection{The hierarchical structure of graph $ \mathcal{G_B} $} 
\begin{figure}[h!] 
	\includegraphics[width=\textwidth]{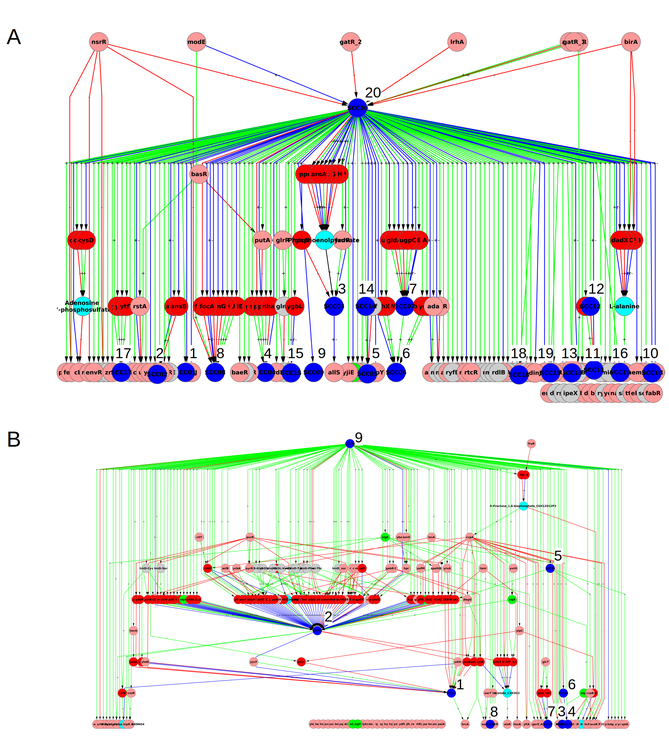}
	\caption[Hierarchical structure of GRNs of \textit{E. coli} and \textit{B. subtilis} with feedbacks from respective metabolic networks: Condensed version of graph $ \mathcal{G_B} $]{\textbf{Hierarchical structure of GRNs of \textit{E. coli} (A) and \textit{B. subtilis} (B) with feedbacks from respective metabolic networks: Condensed version of graph $ \mathcal{G_B} $.} The inclusion of feedbacks from metabolic network into respective GRNs reshuffles the hierarchy, without feedbacks ($ \mathcal{G_A} $), and also resolves the hierarchy further into more levels. The blue nodes, representing SCCs, are numbered and the detail of each numbered SCC is shown in Fig.  \ref{fig:graphB_sccs_ec_bs}. Note that cyan nodes, coding for metabolites are different from blue nodes, representing SCCs. Nodes and edges follow the same colour code as Fig. \ref{fig:graphB_sccs_ec_bs}.}
	\label{fig:graphB_hierarchical_structure_ec_bs}
\end{figure}

With the inclusion of feedbacks into the GRNs of \textit{E. coli} and \textit{B. subtilis} from their associated metabolic networks, the enzyme coding genes and metabolites also get incorporated into the hierarchical structure, both as constituents of new SCCs and also singly. The levels in the hierarchy also get restructured.
E.g., consider the gene \textit{melR}. \textit{melR} occurred in the level 1 of the hierarchy when the feedback from metabolism into the GRN were not considered and did not form a SCC, (visible upon zooming into Fig. \ref{fig:graphA_all}B for \textit{E. coli}). In graph $ \mathcal{G_B} $ it not only forms a SCC (\textit{SCC14}, Fig. \ref{fig:graphB_sccs_ec_bs}A) consisting of four nodes---1 gene coding for TF (\textit{melR}), 2 genes coding for enzymes (\textit{melA}, \textit{melB}) and 1 metabolite (Melibiose)---but also shifts to level 2 (Fig. \ref{fig:graphB_hierarchical_structure_ec_bs}A). While in $ \mathcal{G_A} $ it regulated only genes coding for enzymes, in graph $ \mathcal{G_B} $ it regulates SCC5 having 2 genes coding for TFs, 5 for enzymes, and 1 metabolite (see Fig. \ref{fig:graphB_sccs_ec_bs}A and \ref{fig:graphB_hierarchical_structure_ec_bs}A). 

For \textit{E. coli} graph $ \mathcal{G_B} $ the largest SCC (\textit{SCC20}, Fig. \ref{fig:graphB_hierarchical_structure_ec_bs}A and \ref{fig:graphB_sccs_ec_bs}A) is built by addition of further nodes and links to the largest SCC (\textit{SCC7}, Fig. \ref{fig:graphA_all}B,C) of \textit{E. coli} graph $ \mathcal{G_A} $, and is similarly located at top but one level of the hierarchical organization. For \textit{B. subtilis} graph $ \mathcal{G_B} $, the SCC at the top (\textit{SCC9}, Fig. \ref{fig:graphB_hierarchical_structure_ec_bs}B and \ref{fig:graphB_sccs_ec_bs}A) is same as the one located at the top in case of graph $ \mathcal{G_A} $ (\textit{SCC4}, \ref{fig:graphA_all}B), but is not the largest SCC. The largest SCC for the \textit{B. subtilis} graph $ \mathcal{G_B} $ is \textit{SCC2} which is located at level 4, and is heavily regulated (high in-degree), but regulates few (low out-degree), Fig. \ref{fig:graphB_hierarchical_structure_ec_bs}B and \ref{fig:graphB_sccs_ec_bs}B. This difference exists because amino acid related SCCs do not appear to interact with the SCC containing sigma factors.
This in turn was because 4 out of 6 genes of the largest SCC of \textit{B. subtilis} graph $ \mathcal{G_A} $ coded for $ \sigma $-factors which generally are not modulated by the metabolites, and the other 2  genes coded for TFs received no feedback from metabolism.
It would be interesting to see whether, as the database for \textit{B. subtilis} expands, the amino acid metabolism cluster \textit{SCC2} merges with the housekeeping $ \sigma $-factor cluster \textit{SCC9} (as is the case in \textit{E. coli} \textit{SCC20} in Fig. \ref{fig:graphB_sccs_ec_bs}), or whether they remain separate.


\section{Modules in GRN augmented with functional feedbacks from MN}\label{sec:chap4_identifying_function_modules}
The augmentation of the GRN with feedbacks from the MN---graph $ \mathcal{G_B} $---substantially increased the amount of feedbacks in it (compare Fig. \ref{fig:graphB_sccs_ec_bs} and \ref{fig:graphA_all}C).
In order to understand the structure and role of these feedbacks more clearly we construct another graph, $ \mathcal{G_C} $, which is a sparser version of $ \mathcal{G_B} $.
We use the knowledge that not all but only a part of the metabolic network is active under a given environmental condition (EC) to simplify the augmented GRN by considering only essential feedbacks under a set of defined ECs. This approach allows us to study the functioning of the SCCs with regard to the given environment.
In order to simulate various ECs, we use the computational technique of flux balance analysis (FBA) \cite{varma_metabolic_1994,orth_what_2010} (see Methods \ref{methods_gc_construction} and \ref{methods_fba}). 
The ECs we consider are minimal media characterized by a single organic source of carbon and a few other essential metabolites.
For the present metabolic models, we obtain a list of 158 ECs for \textit{E. coli} and 118 for \textit{B. subtilis}, in which the organism has a positive growth rate (see SI S3 for details).
We determine essential reactions under each given EC, and consider the feedbacks into the GRN from the metabolites taking part in these reactions only.
Next, we augment the GRN with these feedbacks from the MN. We refer to this augmented version of the GRN as graph $ \mathcal{G_C} $ (details in Methods section \ref{methods_gc_construction}; detailed data of $ \mathcal{G_C} $ in SI S4; summary in Table \ref{table:grn_and_mn_numbers_graph_C}). 


A comparison of Table \ref{table:grn_and_mn_numbers_graph_C} with Table \ref{table:grn_and_mn_numbers} shows that the number of metabolite nodes eliminated from \textit{E. coli} and \textit{B. subtilis} networks is 15 and 5 respectively, while the number of links eliminated is 250 and 208 respectively. The procedure (of going from $ \mathcal{G_B} $ to $ \mathcal{G_C} $) eliminates a substantial number of links. Thus while $ \mathcal{G_B} $ and $ \mathcal{G_C} $ both incorporate feedbacks from the metabolic network into the GRN, we can say that $ \mathcal{G_C} $ includes only ``functionally relevant'' feedbacks, since it includes only those reactions and metabolites of $ \mathcal{G_B} $ that are essential for the growth of the organism in some medium.
Next, we analyze the logic of each SCC of $ \mathcal{G_C} $ in detail in terms of its potential influence on the dynamics of the system by considering the signs of the links, the embedding of the SCC in the larger network, and the conditions in which the metabolic reaction link in the SCC is found to be active. This analysis shows that almost every SCC has a specific functional role in the organism and can be associated with a functional module of the system.

\begin{table}[htbp]	
	\centering
	\begin{tabular}{lcc}
		\toprule
		Feature & \textit{E. coli} & \textit{B. subtilis} \\ 
		\midrule 
		Nodes & 3328 & 1705 \\ 
		Genes & 3277 & 1681 \\ 
		Metabolites & 51 & 24 \\ 
		Total edges & 9029 & 3338 \\ 
		Gene to gene edges & 8740 & 3096\\
		Gene to metabolite edges & 229 & 217\\
		Metabolite to gene edges & 60 & 25\\
		Levels in Hierarchy & 12 & 17 \\ 
		Number of SCCs & 28 & 14 \\ 
		Size of largest SCC & 97 & 13 \\  
		\bottomrule
	\end{tabular}
	\vspace{.5cm}
	\caption[GRN augmented with functional feedbacks from metabolic network.]{\textbf{GRN augmented with functional feedbacks from metabolic network ($ \mathcal{G_C} $).} Details of composition of GRN augmented with feedbacks from metabolic network---graph $ \mathcal{G_C} $---for \textit{E. coli} and \textit{B. subtilis}.}
	\label{table:grn_and_mn_numbers_graph_C}
\end{table}

\subsection{Strongly connected components and hierarchical structure of graph $ \mathcal{G_C} $}
\begin{figure}[h!] 
	\includegraphics[width=.8\textwidth]{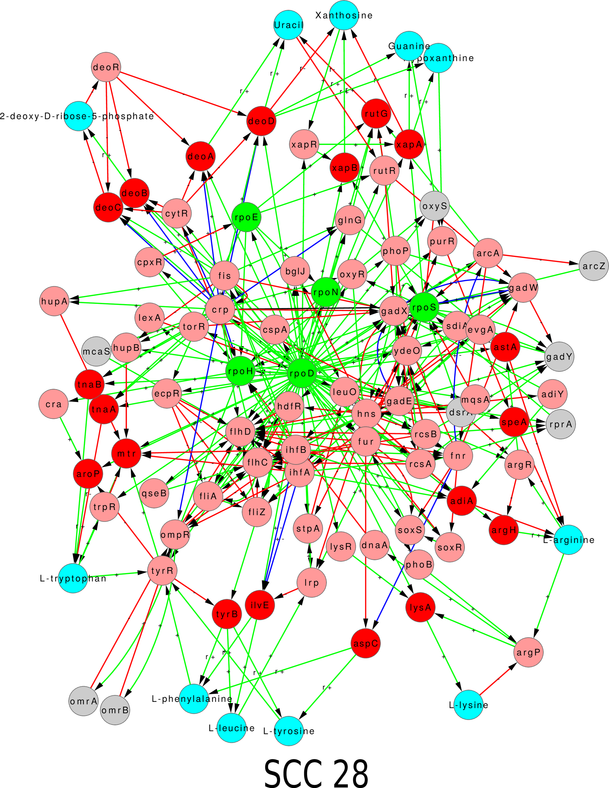}
	\caption[LSCC of GRN of \textit{E. coli} graph $ \mathcal{G_C} $.]{\textbf{Largest strongly connected component of \textit{E. coli} graph $ \mathcal{G_C} $ (GRN augmented with functionally relevant feedbacks from metabolic network).} Nodes and edges follow the same colour code as in Fig. \ref{fig:graphB_sccs_ec_bs}. }
	\label{fig:graphC_lscc_ess_ec}
\end{figure}

\begin{figure}[h!] 
	\includegraphics[width=\textwidth]{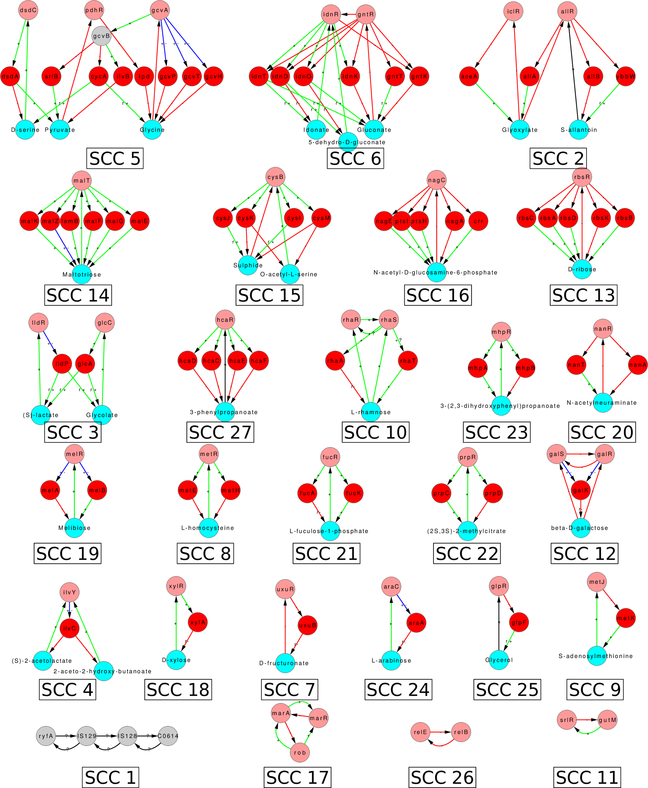}
	\caption[Modules or kernel sub-systems of GRN of \textit{E. coli} graph $ \mathcal{G_C} $.]{\textbf{Smaller strongly connected components of \textit{E. coli} graph $ \mathcal{G_C} $.} Nodes and edges follow the same colour code as in Fig. \ref{fig:graphB_sccs_ec_bs}.}
	\label{fig:graphC_ssccs_ess_ec}
\end{figure}

\begin{figure}[h!] 
	\includegraphics[width=\textwidth]{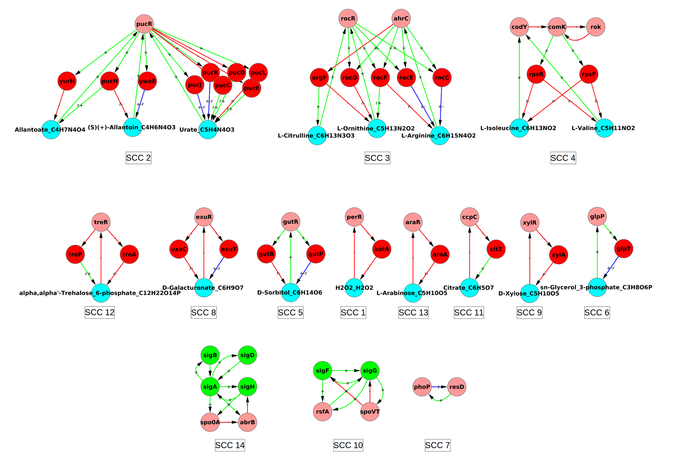}
	\caption[Strongly connected components of GRN of \textit{B. subtilis} graph $ \mathcal{G_C} $.]{\textbf{Strongly connected components of GRN of \textit{B. subtilis} graph $ \mathcal{G_C} $.} The resulting SCCs of \textit{B. subtilis} graph $ \mathcal{G_C} $ are enumerated. Nodes and edges follow the same colour code as in Fig. \ref{fig:graphB_hierarchical_structure_ec_bs}}
	\label{fig:graphC_sccs_ess_bs}
\end{figure}

\begin{figure}[h!] 
	\includegraphics[width=\textwidth]{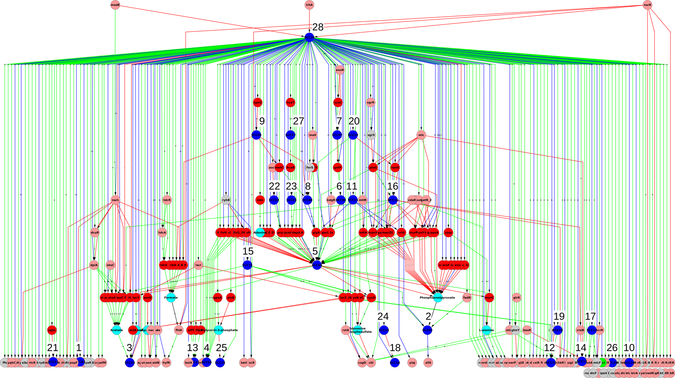}
	\caption[The hierarchical structure of GRN of \textit{E. coli}: Condensed version of graph $ \mathcal{G_C} $]{\textbf{The hierarchical structure of GRN of \textit{E. coli}: Condensed version graph $ \mathcal{G_C} $.} The numbered blue nodes are SCCs whose detail is shown in Figs. \ref{fig:graphC_lscc_ess_ec} and \ref{fig:graphC_ssccs_ess_ec}. Nodes and edges follow the same colour code as in Fig. \ref{fig:graphB_sccs_ec_bs}.}
	\label{fig:graphC_hierarchical_ess_ec}
\end{figure}

\begin{figure}[h!] 
	\includegraphics[width=\textwidth]{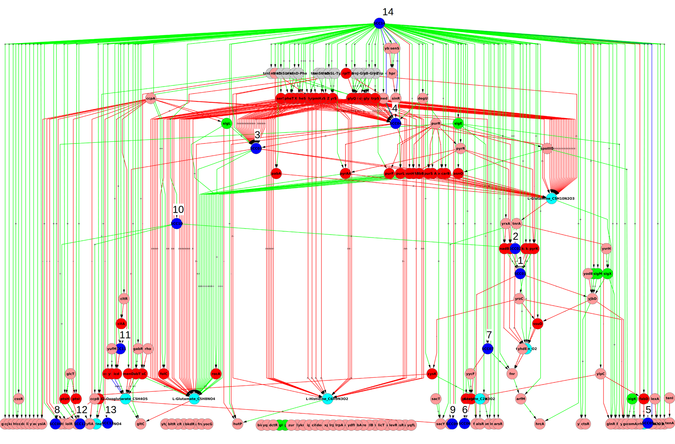}
	\caption[The hierarchical structure of GRN of \textit{B. subtilis}: Condensed version of graph $ \mathcal{G_C} $]{\textbf{The hierarchical structure of GRN of \textit{B. subtilis}: Condensed version of graph $ \mathcal{G_C} $.} The numbered blue nodes are SCCs whose detail is shown in Fig. \ref{fig:graphC_sccs_ess_bs}. Nodes and edges follow the same color code as in Fig. \ref{fig:graphB_sccs_ec_bs}.}
	\label{fig:graphC_hierarchical_ess_bs}
\end{figure}

The SCCs and condensed graph of $ \mathcal{G_C} $ for \textit{E. coli} are given in Figs. \ref{fig:graphC_lscc_ess_ec}, \ref{fig:graphC_ssccs_ess_ec} and \ref{fig:graphC_hierarchical_ess_ec}, and those of \textit{B. subtilis} in Figs. \ref{fig:graphC_sccs_ess_bs} and \ref{fig:graphC_hierarchical_ess_bs}. 
The data of graph $ G_C $ for both organisms, is given in SI S4, including nodes, links, distribution of nodes in hierarchical levels, details of each SCC, etc.
There are more SCCs in graph $ \mathcal{G_C} $ than in graph $ \mathcal{G_B} $ (compare the numbers in Tables \ref{table:grn_and_mn_numbers} and \ref{table:grn_and_mn_numbers_graph_C}).
The size of largest SCC of graph $ \mathcal{G_C} $ for \textit{E. coli} is considerably reduced (SCC 28 in Fig. \ref{fig:graphC_lscc_ess_ec} has 97 nodes as compared to SCC 20 in Fig. \ref{fig:graphB_sccs_ec_bs} having 378 nodes). Many sub-networks of the largest SCC of $ \mathcal{G_B} $ now break out into individual smaller SCCs of $ \mathcal{G_C} $ 
(SCCs 3, 5, 6, 7, 8, 9, 11, 14, 15, 16, 17, 20, 22, 23, and 27 in Fig. \ref{fig:graphC_ssccs_ess_ec}) 
suggesting that the latter modules function independently in the conditions considered here. Some of the SCCs of $ \mathcal{G_B} $ (SCC 2, 4, 6, 8, 10, 16, and 18) are absent in $\mathcal{G_C}$ because the corresponding metabolite is not a minimal medium food source in our considered ECs. 
For \textit{B. subtilis} a similar phenomenon happens and the largest SCC of $ \mathcal{G_C} $ is of size only 13. The LSCC of graph $ \mathcal{G_B} $ (size 85; SCC 2 in Fig. \ref{fig:graphB_sccs_ec_bs}B) breaks up into smaller SCCs (SCC 1, 2, 3, 4 and 11  in Fig.
\ref{fig:graphC_sccs_ess_bs}) of its constituent communities of allantoin, amino acids and citrate modules. The number of levels in the condensed graph of $ \mathcal{G_C} $ for \textit{E. coli} and \textit{B. subtilis} are 12 and 17 respectively, Figs. \ref{fig:graphC_hierarchical_ess_ec}, \ref{fig:graphC_hierarchical_ess_bs}. 

\subsection{Architecture and functionality of modules}\label{sec:chap4_architecture_and_functionality_of_modules}
The question arises as to what the significance is of the non-trivial SCCs. The dynamics of an SCC by definition depends upon the state of the nodes above it in the GRN hierarchy and on its own non-trivial internal structure. To study their functional significance we examine the internal structure of the SCCs obtained above for \textit{E. coli} and \textit{B. subtilis} in detail. We show below that almost every SCC obtained above is associated with an identifiable and definite functionality, or that it performs a specific task or set of tasks in the organism. It may be noted that a priori, it is not obvious that a non-trivial SCC should have a specific biological role. The fact that we find most SCCs to have a specific role suggests that this is a useful construct. The twin properties of being dynamically relatively autonomous and having an identifiable biological functionality justify our referring to these SCCs as associated with `modules' of the GRN. Many algorithmic methods exist in the literature of identifying modules, e.g., those that cluster expression data and those that identify `communities' in networks. Our algorithmic method is different, and it does not identify all modules in the system, but those that it does have a fairly tight functional significance, as will be seen below. Thus the method would be particularly useful in identifying modules for those organisms for which the genetic and metabolic databases will become available, but whose biology has not been as extensively studied as \textit{E. coli} and \textit{B. subtilis}.

The SCCs of Figs. \ref{fig:graphC_ssccs_ess_ec} and \ref{fig:graphC_sccs_ess_bs} have been displayed along with their properties in SI \nameref{S5_ec_graphC_module_analysis} and \nameref{S6_bs_graphC_module_analysis}. We have focused on SCCs that have at least one metabolite. The total number of such SCCs is 23 in \textit{E. coli} and 11 in \textit{B. subtilis}. For \textit{E. coli} the largest SCC, SCC 28 in Fig. \ref{fig:graphC_lscc_ess_ec}, is complicated, has multiple functional tasks, and is discussed separately in a later subsection.

It is convenient to organize the functional study of the metabolite containing SCCs, hereafter referred to interchangeably as modules, in ascending order of size (number of nodes), as the functionality of larger modules can often be understood in terms of smaller structures. 

\subsubsection{Size 3 modules: NFL or PFL}\label{sec:chap4_size_3_modules}
\begin{figure}[h!] 
	\includegraphics[width=0.95\textwidth,trim={0 2in 0 0},clip]{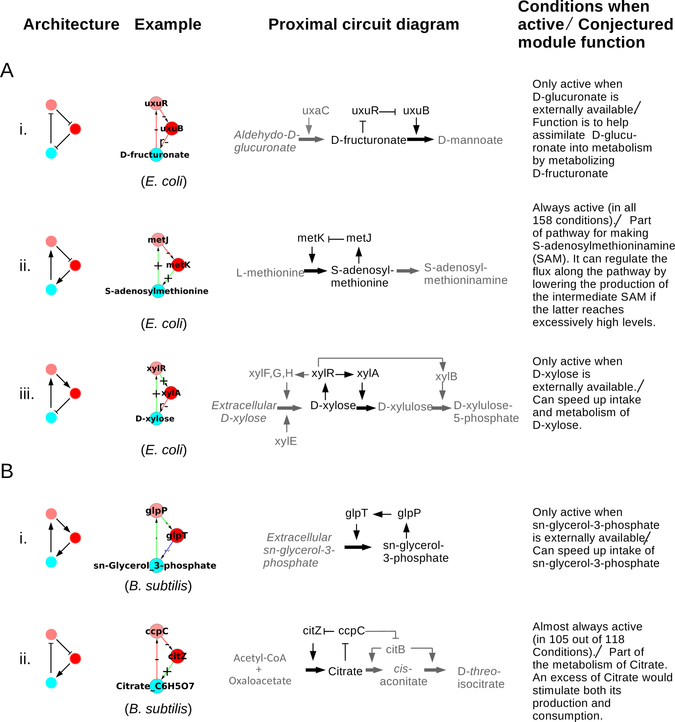}
	\caption[Architecture and function of 3-node modules.]{\textbf{Architecture and function of 3-node modules.} \textbf{A.} The negative feedback loop (NFL). \textbf{B.} The positive feedback loop (PFL). Beside the architecture an example of the same is shown, as well its proximal circuit diagram showing the elementary pathway logic of the involved genes and metabolites. In the latter diagram the genes and metabolites of the example module are shown in black, while the preceding and following elements of the pathway have been greyed, for distinction and clarity. Thick arrows represent metabolic reactions or conversion of one metabolite into another. An arrow from an enzyme coding gene to a thick arrow means that the enzyme catalyzes the reaction. Other arrows follow the same convention as in Figs. \ref{fig:schematic_link_interpretation_and_grn_feedback_from_mn} and \ref{fig:graphB_sccs_ec_bs}. A flat tipped link from a metabolite to a TF means that when the metabolite binds to the TF it inactivates it, i.e., prevents it from binding to its target gene. We will say that an SCC is `active' in an EC if a reaction (enzyme gene to metabolite link) belonging to the SCC is essential in that EC. The last column gives the environmental conditions in which the SCC is active, and the conjectured functional role of the module.}
	\label{fig:graphC_architecture_3node_modules}
	
\end{figure}

There are five size 3 modules in \textit{E. coli} and five in \textit{B. subtilis}. Each size 3 module has one metabolite, one TF gene and one enzyme gene. 
The modules of size 3 can be classified into two categories: negative feedback loop and positive feedback loop (NFL, PFL).
While the composite logic in these modules remain either a NFL or PFL, their internal composition in terms of position and sign of the interaction may vary.
Fig. \ref{fig:graphC_architecture_3node_modules} tabulates examples of this structural variation that we find and details about each example displayed that leads to a statement about the possible functionality of the module. 
All the modules are shown in SI \nameref{S5_ec_graphC_module_analysis} and \nameref{S6_bs_graphC_module_analysis}. As an example, we discuss the module in the first row of Fig. \ref{fig:graphC_architecture_3node_modules}. This module is found to be active only when the food source is Aldehydo-D-glucoronate. In that situation enzyme \textit{UxaC} converts this molecule to D-fructuronate. One can guess that the function of the module, a NFL with three inhibitory links, is to metabolize D-fructuronate when it is in excess into a further downstream product. The internal structure of the module shows how this can be achieved. D-fructuronate binds to the TF \textit{UxuR} and inactivates it. \textit{UxuR} was repressing the expression of gene \textit{uxuB} and hence preventing the production of enzyme \textit{UxuB}. When \textit{UxuR} is inactivated by D-fructuronate, the enzyme \textit{UxuB} is produced and this in turn catalyzes a reaction that consumes D-fructuronate. This functionality is useful when D-fructuronate is in excess, consistent with the fact that we only see it when aldehydo-D-glucoronate is the food source. It helps the assimilation of this food source into the metabolism. Further it prevents gene expression (which is costly \cite{dekel_alon_optimality_2005}) of \textit{uxuB} when not required, i.e., when the food source is not available.

Almost every topology we have found as a module has a clearly identifiable functionality or purpose which is evident from Fig. \ref{fig:graphC_architecture_3node_modules}. Broadly, two functions are served. (1) The modules of D-glucoronate, D-xylose and sn-Glycerol-3-phosphate---Fig. \ref{fig:graphC_architecture_3node_modules}A(i and iii), B(i) respectively---are only active when the respective metabolite is itself the food source as the growth media, else they are inactive. These modules are present at the input end of the metabolic network. Their specific activity indicates that they function for the uptake or metabolism of respective metabolite food molecule in the cell. (2) The modules which are active in all or almost all of the growth conditions and are located deep in some pathway of the metabolite, e.g. the modules of S-adenosylmethionine and citrate, Fig.\ref{fig:graphC_architecture_3node_modules}A(ii), B(ii). They can have different functions depending upon the structure. E.g., the S-adenosylmethionine module, Fig. \ref{fig:graphC_architecture_3node_modules}A(ii), seems to be designed to regulate the production (over/under- production) of S-adenosylmethionine, whereas the citrate module, Fig. \ref{fig:graphC_architecture_3node_modules}B(ii), seems to be designed to speed up the production of citrate. We note that out of the ten size-3 modules, four belong to category A(i) and one each to A(ii), A(iii), B(i) and B(ii). Of the remaining two, in one case the link from the TF to the enzyme is stated to have a dual sign (+ or -) in the database, and for the other there is some ambiguity about the existence of the link from the metabolite to the TF. Details are in the SI \nameref{S5_ec_graphC_module_analysis} and \nameref{S6_bs_graphC_module_analysis}.

It is important to point out that the SCCs we find often uncover a larger module of which they are a part. The larger module can be constructed once the SCC is identified by our method. An example is the D-xylose module in Fig \ref{fig:graphC_architecture_3node_modules}A(iii). The SCC found by our method only had the 3 nodes: the metabolite D-xylose, the TF \textit{XylR} and the enzyme \textit{XylA}. However the proximal circuit shows that \textit{XylR} also activates enzymes \textit{XylF,G,H} which catalyze the intake of extracellular D-xylose and \textit{XylB} which converts D-xylulose into D-xylulose-5-phosphate. The \textit{XylF,G,H} do not appear as a module in our procedure because its intake reaction is outside the scope our procedure as it is not an essential reaction because another enzyme \textit{XylE} provides a second pathway for the intake of extracellular D-xylose. Collectively this whole system should be considered the ``D-xylose module" instead of the 3 node SCC found by our method. 
Thus our method sometimes detects not the whole module but what might be called as `kernel' of the module from which the rest of the module can often be constructed by examining the proximal circuit.
We could have discovered that \textit{XylB} is also part of the SCC if we had been more liberal in including metabolic reactions that produce or consume not just metabolites that bind to TFs but also their nearest neighbour metabolites. It is a task for the future to extend our methodology to include these effects, and to see the usefulness of the results so obtained.

\begin{figure}[h] 
	\centering
	\includegraphics[width=.5\textwidth]{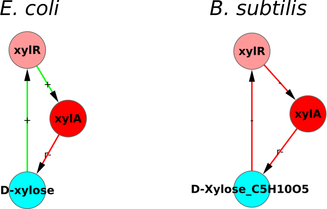}
	\caption[The size-3 D-xylose module (NFL) with same function but different internal structure.]{\textbf{The size-3 D-xylose module (NFL) with same function but different internal structure.} The D-xylose module in both \textit{E. coli} and \textit{B. subtilis} is an over-all negative feedback loop serving the same function of intake and metabolism of D-xylose, but has different internal structures in the two organisms. This is an example of the conservation of function of a module despite variations in its internal structure.}
	\label{fig:graphC_xylose}
\end{figure}

The D-xylose module also illustrates another interesting feature: the overall function is conserved across organisms in spite of the variation in the internal structure of the module. D-xylose forms a size-3 SCC in both \textit{E. coli} and \textit{B. subtilis}, Fig. \ref{fig:graphC_xylose}. In both cases the module is a negative feedback loop (NFL). Also, in both cases, the module is active only when D-xylose is the food source and can speed up its intake and metabolism. However, the D-xylose module in \textit{E. coli} has a different internal structure than that in \textit{B. subtilis}. In \textit{E. coli} the structure of the loop is \textit{XylR} $ \rightarrow $ \textit{XylA} \inactivates{} D-xylose $ \rightarrow $ \textit{XylR} (an overall NFL). Whereas in \textit{B. subtilis} the structure is \textit{XylR} 
\inactivates{}  \textit{XylA}
\inactivates{} D-xylose \inactivates{} \textit{XylR} (also an overall NFL). In \textit{E. coli} the D-xylose activates the TF \textit{XylR} whose action is activation of enzyme \textit{XylA} thus leading to the metabolism of D-xylose into Xylulose. Whereas in \textit{B. subtilis} D-xylose is also metabolized into Xylulose by the enzyme \textit{XylA} but through a different regulation: D-xylose inactivates the TF \textit{XylR} whose action is repression of the enzyme \textit{XylA}, with the overall effect the enzyme \textit{XylA} again metabolizing D-xylose to Xylulose. 
Different ways of implementing the same logic can have similar effects in a steady state but could differ in transient effects.

\subsubsection{Size 4 modules: consumer motifs} 
The modules of size 4 can be broadly divided into two classes: (a) a 4-node motif associated with uptake and utilization of small molecules, (b) two essentially 3-node motifs sharing nodes and therefore forming a 4-node SCC. In the former case the full  4-node structure is essential to understand its functionality. In the latter case the individual 3-node sub-structures have a functionality in their own right.

\begin{figure}[h!] 
	\includegraphics[width=\textwidth]{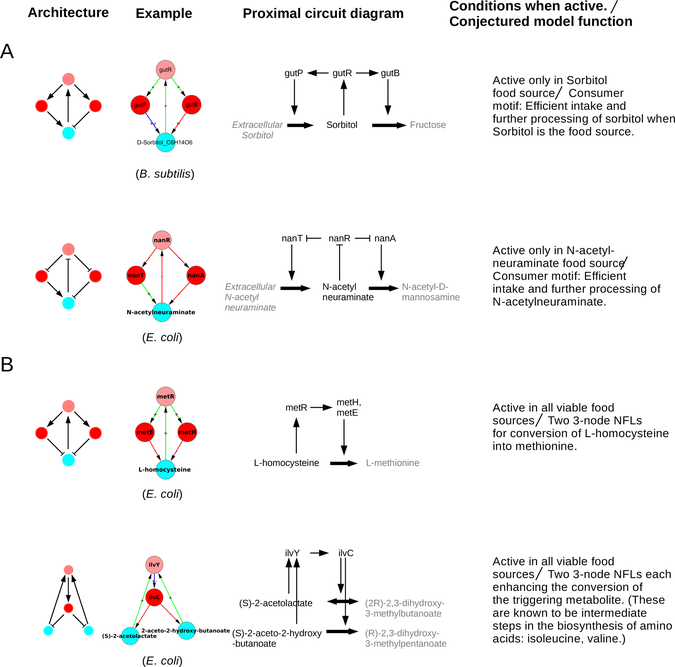}
	\caption[Architecture of 4-node modules.]{\textbf{Architecture and function of 4-node modules.} \textbf{A.} The consumer motif along with its two identified variations. \textbf{B.} The 4-node modules which can be reduced to essentially two 3-node NFLs with shared nodes. Beside the architecture an example of the same is shown, as well as the elementary pathway logic of the genes and metabolites involved in the module, and functionality of the structure. Nomenclature is as in Fig. \ref{fig:graphC_architecture_3node_modules}.}
	\label{fig:graphC_architecture_4node_modules}
\end{figure}

The first class (Fig. \ref{fig:graphC_architecture_4node_modules}A) is a structure suitable for the uptake and metabolism of food molecule, studied in \cite{krishna_combinatorics_2007} and referred to as a `consumer motif'. The consumer motif can be thought of as a combination of two 3-node motifs---PFL followed by NFL---where both the loops work in tandem for the uptake and metabolism of the food molecule. In a consumer motif the first loop serves to transport the metabolite into the cell from the outside, i.e., increase its cellular concentration, while the second loop metabolizes it to other compounds thereby decreasing its cellular concentration. We found that the majority of the 4-node motifs (6 out of 8 in \textit{E. coli} and all 3 in \textit{B. subtilis}, hence 9 out of a total of 11 in both organisms) are of consumer type. Example of both the variants of the consumer motif, one where the metabolite activates the TF (the sorbitol module) and the other where the metabolite inactivates the TF (the N-acetylneuraminate module), are shown in Fig. \ref{fig:graphC_architecture_4node_modules}A. We note that in most cases the consumer motif is located at the uptake end of a metabolite's pathway---6 out of 9 are involved in uptake and further processing of a metabolite. Of the remaining three, two are not involved in uptake but are close enough to the uptake reaction for their role to be reasonably clear (the dihydroxyphenol propanoate module and fuculose module). All these 8 modules are active with only one or a few very specific food sources. One (the methylcitrate module) is not located immediately following external metabolite. Again its function is clear from the structure, but the reason for its deep location in the network is not clear (see SI \nameref{S5_ec_graphC_module_analysis} and \nameref{S6_bs_graphC_module_analysis} for details).

In the second class (Fig. \ref{fig:graphC_architecture_4node_modules}B), the functionally non-reducible modules involved are the 3-node motifs. The only two cases in this class, the homocysteine module and the acetolactate-acetohydroxybutanoate module are both shown in Fig. \ref{fig:graphC_architecture_4node_modules}B. These modules are active (present in) all the food source growth conditions where they are part of the reaction pathways for synthesis of amino acids, here methionine, isoleucine and valine. The elementary pathway logic for these modules suggests that they perform the familiar NFL functions of homeostasis. 

\subsubsection{Size $ \geq5 $ modules}
As the size of modules increases to 5 and beyond, the complexity of the modules also increases. 
Several of these modules contain more than one metabolite binding to TFs, and some of them contain both more than one metabolite and TF. They also have `sub-modules' which are active in different subsets of conditions. Nevertheless, an analysis of their graphs including the sign of the links, their proximal circuit diagram, and conditions in which they are active, allows us to guess their functional roles in most cases without going into further biological details. There are 10 modules of size $ \geq $5 in \textit{E. coli} graph $ \mathcal{G_C} $ and 3 in \textit{B. subtilis}, giving a total of 13 modules. Of these 13, one is a spurious module in \textit{B. subtilis} (that arises because FBA introduces some fictitious reactions in the metabolic network corresponding to biomass production, see detail in SI \nameref{S6_bs_graphC_module_analysis}) and should be ignored. 
Of the remaining 12 we could identify functional roles quite clearly for 8 (all in \textit{E. coli}), partially for 2 (both in \textit{B. subtilis}) and we were unable to do so for 2 of the modules (both in \textit{E. coli}). Most (9 out of 12) of the modules of size $ \geq $5 were just more complex versions of the 4-node consumer motifs and the 3-node feedback loops or their combinations. The details are given in SI \nameref{S5_ec_graphC_module_analysis} and \nameref{S6_bs_graphC_module_analysis}. 

\begin{figure}[h!] %
	\vspace*{-1cm}
	\centering
	\includegraphics[width=0.85\textwidth]{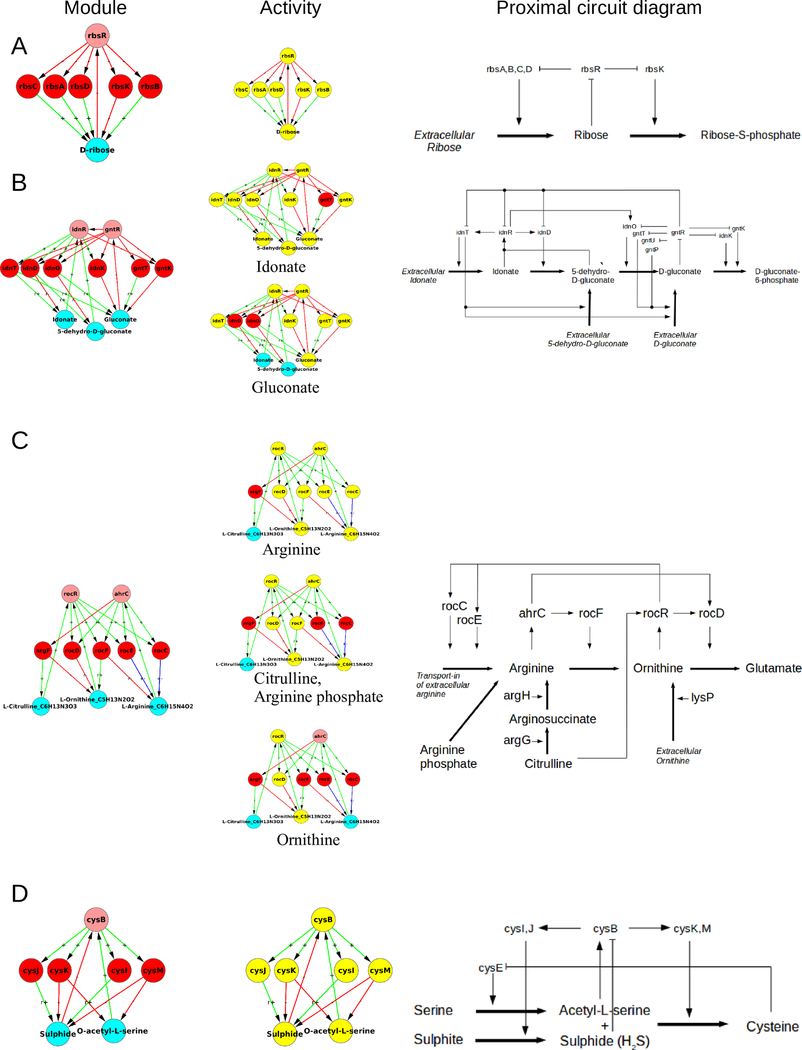}
	\caption[Modules of higher size ($ \ge5 $)]{\textbf{Modules of higher size ($ \ge5 $)} A gist of modules of size $ \ge5 $ of various types is shown through examples in the figure. The three columns show the module, the active subgraph (yellow nodes) of the module under the indicated carbon food source, and its proximal circuit diagram. The food source defines a minimal growth medium (environmental condition, EC). By `active subgraph' we mean the SCC formed when only the reactions essential in that condition are taken into account and the non-essential reactions in that condition are excluded.
	\textbf{(A)} The Ribose module showing an example of a multigene enzyme, a trivial extension of a 4-node consumer motif. 
	\textbf{(B)} The Idonate-Gluconate module showing an example of sequentially nested consumer motifs---a structure that effectively metabolizes multiple food sources located on a single pathway in the metabolic network.
	\textbf{(C)} The Citrulline-Ornithine-Arginine module showing  an example of a module whose different subgrahs are active under different conditions. \textbf{(D)} The Cysteine module, a new multinode architecture. Nomenclature of nodes is as in Fig. \ref{fig:graphC_architecture_3node_modules}.}
	\label{fig:architecture_module_ge5}.
\end{figure}

A gist of modules of size $\geq 5$ is shown in Fig. \ref{fig:architecture_module_ge5}. Broadly, three categories emerge: (1) Modules where multiple genes contribute to the formation of an enzyme, e.g., the Ribose module, (Fig. \ref{fig:architecture_module_ge5}A) and Phenylpropanoate module (given in SI \nameref{S5_ec_graphC_module_analysis}). In the Ribose module (Fig. \ref{fig:architecture_module_ge5}A), the genes \textit{rbsA}, \textit{rbsB} and \textit{rbsC} contribute to the formation of the ribose-ABC transporter. The \textit{RbsD} protein functions for efficient uptake of ribose when it is transported into the cell via a mutated form of glucose transporter (\textit{PtsG}). This 7 node motif is thus essentially like a 4-node consumer motif in which 4 of the enzyme coding genes are performing the same function of transporting the metabolite into the cell.
(2) Multiple simple motifs joined together, e.g., Idonate-Gluconate module (Fig. \ref{fig:architecture_module_ge5}B). 
This is a set of three mutually reinforcing consumer motifs. There is a metabolic pathway of conversion of idonate to 5-dehydro-D-gluconate to D-gluconate to D-gluconate-6-phosphate in the cell. The first three metabolites can also be external food sources. Inspection of the proximal circuit diagram and the nodes highlighted in yellow in Fig. \ref{fig:architecture_module_ge5}B reveals that when D-gluconate is the food source this circuit will switch on the corresponding consumer motif to produce the last molecule in the pathway, while the part of the circuit before D-gluconate is switched off. Similarly, when Idonate is the food source, the enzymes corresponding to its uptake and metabolism are active (highlighted in yellow) while transporters of the other intermediate food sources (D-gluconate, 5-dehydro-D-gluconate) are off. As another example of the situation in which different parts of the module are active in different conditions we show the Arginine-Ornithine-Citrulline module in Fig. \ref{fig:architecture_module_ge5}C. This also decomposes into simpler motifs active in different conditions. Some nodes and links of the module are shared by parts active in different conditions; these are thereby multitasking and contributing to the overall economy of the structure. Here again the structure of the different active parts is essentially one that has already been encountered earlier in size 3 and 4 modules.
(3) Modules which are qualitatively different in structure and dynamics from those of size 3 and 4. An example of this is the Sulphide-Acetylserine module (or Cysteine module) (Fig. \ref{fig:architecture_module_ge5}D), discussed below. 

The following logic can be ascribed to the Sulphide-Acetylserine (or Cysteine) module for the controlled production of Cysteine, assuming that Serine and Sulphite are available in the cell. If Cysteine levels are sufficiently high, it inhibits the enzyme \textit{CysE} that catalyzes the conversion of Serine to Acetyl-L-serine, thereby blocking its own production. When Cysteine level falls sufficiently low, the inhibition of \textit{CysE} is lifted, allowing for the production of Acetyl-L-serine. This activates \textit{CysB} which in turn activates \textit{cysI,J} leading to the production of Sulphide from Sulphite. \textit{CysB} also activates \textit{cysK,M} which catalyze the reaction between Acetyl-L-serine and Sulphide (both now available as reactants) to produce Cysteine. Over-production of Cysteine shuts the module off by inhibition of \textit{CysE}. Sulphide at high level also represses \textit{CysB} to control its own over-production. This module is well known in the biological literature for the regulation of Cysteine production \cite{kredich_molecular_1992}. This supports the claim that our algorithmic and blind approach starting from the databases is capable of retrieving biologically relevant functional modules at the local level in the network, while also providing information about the global organization of the modules in the network.

We emphasize that our procedure of identifying the SCCs is blind to the signs of edges between the nodes. Still the signs in the modules we obtain through our procedure conform very well to the intuitive logical functioning of these modules. This suggests thats our approach which employs construct of SCC is actually able to find out functionally relevant biological modules or dynamical systems.


The two modules that defied a proper classification, namely, the Serine-Pyruvate-Glycine module and the Allantoin-Glyoxylate module are also more complex (both in \textit{E. coli} and active in 3 and 10 conditions respectively; see SI S5).

\section{Structure of core} \label{sec:chap4_structure_of_core}
Consider the largest SCC (LSCC) in \textit{E. coli} graph $ \mathcal{G_C}$ (Fig. \ref{fig:graphC_lscc_ess_ec}) and its location in $ \mathcal{G_C}$ (Fig. \ref{fig:graphC_hierarchical_ess_ec}). We can call the LSCC as the `core' of the network as it contains the largest number of feedbacks and is positioned at the top in the hierarchical structure of the network, and hence has the capacity to influence the whole of the downstream GRN. The LSCC or core of the \textit{E. coli} graph $ \mathcal{G_C}$ comprises of a total of 97 nodes (5 $ \sigma $-factors, 54 TFs, 8 non-coding RNAs, 19 enzyme-genes and 11 metabolites), Fig. \ref{fig:graphC_lscc_ess_ec}. In  case of \textit{B. subtilis}, While there is an SCC at the top of the hierarchy (Fig. \ref{fig:graphC_hierarchical_ess_bs}), it is not the largest SCC (see SCC 14 in Fig. \ref{fig:graphC_sccs_ess_bs}). The SCCs larger than SCC 14 are lower down in the hierarchy and seem to influence a relatively smaller no. of downstream nodes. It is quite possible that this situation is a consequence of the incompleteness of the \textit{B. subtilis} network (in earlier versions of \textit{E. coli} network the SCC at the to of the hierarchy was also small). We therefore discuss the LSCC of \textit{E. coli} only in the remainder of this section. 

The core of $ \mathcal{G_C} $ for \textit{E. coli}, though much smaller than the core of $ \mathcal{G_B} $ (compare SCC 28 of $ \mathcal{G_C} $ in Fig. \ref{fig:graphC_lscc_ess_ec} with SCC 20 of $ \mathcal{G_B} $ in Fig. \ref{fig:graphB_sccs_ec_bs}A) is still quite large and complex and it is difficult to assign a specific functionality to it. In fact it is obvious from the location of SCC 28 in Fig. \ref{fig:graphC_hierarchical_ess_ec} and the fact that links emanating from it go to almost all downstream nodes that the core of $ \mathcal{G_C} $ is involved in regulating almost all functions of the cell. Nevertheless it is worthwhile to ask if the core of $ \mathcal{G_C} $ has a further substructure that is meaningful. 

In this context we take a cue from the $ \mathcal{G_C} $ of \textit{B. subtilis}.
We note that in the case of $\mathcal{G_C}$ of \textit{B. subtilis} the $ \sigma $-factors make  separate SCCs that have no metabolite node (SCC 10 and 14 in Fig. \ref{fig:graphC_sccs_ess_bs}). However, in the case of $ \mathcal{G_C} $ of \textit{E. coli} the $ \sigma $-factors are part of the LSCC (the 5 green nodes in Fig. \ref{fig:graphC_lscc_ess_ec}) along with several metabolite, enzyme and TF nodes. This difference between the SCCs of the $ \mathcal{G_C} $ graphs of \textit{E. coli} and \textit{B. subtilis} suggested the possibility that $ \sigma $-factors in \textit{E. coli} act as a `glue' for connecting different sub-structures of the `core'.
Proceeding on this lead we asked what would happen to the structure of LSCC of graph $ \mathcal{G_C}$ of \textit{E. coli} if we separated the SCC formed by the $ \sigma $-factors from it. This was done by deleting all the incoming links to $ \sigma $-factors from nodes other than the $ \sigma $-factors in the graph shown in Fig. \ref{fig:graphC_lscc_ess_ec}. The number of links deleted was 21. Note that this is a somewhat ad-hoc procedure, but since $ \sigma $-factors typically have relatively broad roles that play out in good growth condition conditions or different types of stress conditions, this may be justified for our purpose of revealing other close relationships in the network.

\begin{figure}[t] 
	\centering
	\includegraphics[width=.8\textwidth]{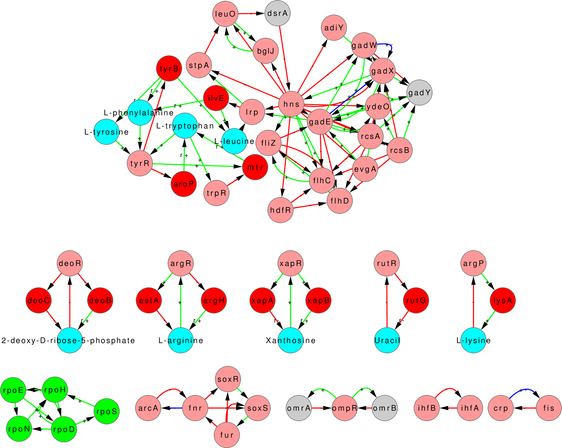}
	\caption[Sub-structure of `core' of \textit{E. coli} graph $ \mathcal{G_C}$]{\textbf{Sub-structure of core of \textit{E. coli} graph $ \mathcal{G_C}$.} The SCCs constituting the sub-structure of the core revealed upon removal of in-links to $ \sigma $-factors from the core (shown in Fig. \ref{fig:graphC_lscc_ess_ec}) of \textit{E. coli} graph $ \mathcal{G_C} $.}
	\label{fig:graphC_core_broken_sccs}
\end{figure}

Taking out the SCC formed by $ \sigma $-factors in this manner broke the core into a total of 11 smaller SCCs (including the $ \sigma $-factor one) the largest of which is of size 29, Fig. \ref{fig:graphC_core_broken_sccs}. 5 of the 11 SCCs are composed of only the gene-nodes, while 6 of the SCCs have at-least one metabolite node. The 6 SCCs containing the metabolites can again be classified into modules of size 3 which here are the  NFLs (the Lysine and Uracil modules); module of size 4 two of which are consumer-motifs (the 2-deoxy-D-ribose-5-phosphate and Xanthosine modules) and one a combination of two NFLs (the L-arginine module). The largest of the module obtained upon the removal of the SCC formed by $ \sigma $-factors, size 29, has predominantly genes coding for TF and involve multiple global regulators with varying functions, which prohibits assignment of a simple function to this module. The functioning of the core remains an open question. 

\section{Discussion}
In this work we have: (i) described the causal structure of GRNs without and with feedbacks from MN, (ii) presented a framework and automated method using graph theory and FBA for determining modules in organisms from a knowledge of their GRN, MN and feedbacks from the MN to GRN, and (iii) applied the method in (ii) to \textit{E. coli} and \textit{B. subtilis} to produce a list of modules in these organisms and discussed the functional role of each module.

\subsection{Causal and computational structures of GRNs}
We employed graph theoretic procedures of strongly connected components (SCCs) and leaf-removal to identify cyclic regions of the GRN and to organize it into a hierarchical structure composed of different levels. 
The hierarchical organization of the GRNs in section \ref{sec:chap3_full_grn} shows that the complicated looking GRNs of \textit{E. coli} and \textit{B. subtilis} can be organized into a causal, largely acyclic architecture, with a few islands of feedback, Fig. \ref{fig:graphA_all}B, consistent with earlier studies. The largest SCC sits near the top of the hierarchy, and is significantly larger and has many more feedbacks than the second largest SCC.

Further, our work takes into account the feedback from metabolic network into the gene regulatory network while illustrating its causal and computational structure, Figs. \ref{fig:graphB_hierarchical_structure_ec_bs}, \ref{fig:graphC_hierarchical_ess_ec} and \ref{fig:graphC_hierarchical_ess_bs}. We find that most feedbacks from the MN to the GRN affect the network only locally while a few feedbacks have a global impact, Fig. \ref{fig:schematic_link_interpretation_and_grn_feedback_from_mn}B,C. Further, some of the qualitative features of the organization (hierarchical structure with mostly small islands of feedback and a large globally regulating SCC near the top of the hierarchy) do not change substantially---witness the similarity of the condensed graphs of $ \mathcal{G_B} $ in Fig. \ref{fig:graphB_hierarchical_structure_ec_bs} to those of $ \mathcal{G_A} $ in Fig. \ref{fig:graphA_all}B.
However, the feedback from the MN substantially increases the complexity of the GRNs of both \textit{E. coli} and \textit{B. subtilis} by increasing the number and sizes of SCCs, especially the size of the largest SCC (Figs. \ref{fig:graphB_sccs_ec_bs}, \ref{fig:graphC_lscc_ess_ec}, \ref{fig:graphC_ssccs_ess_ec} and \ref{fig:graphC_sccs_ess_bs}) and introducing additional levels in the condensed graphs, Figs. \ref{fig:graphB_hierarchical_structure_ec_bs}, \ref{fig:graphC_hierarchical_ess_ec} and \ref{fig:graphC_hierarchical_ess_bs}. 
Further, certain enzymes that were at the lowest level in the hierarchy earlier, along with metabolites move up in the hierarchy above many TFs, thereby changing the causal ordering of the hierarchy.

\subsection{Automated method of determining modules}
We have also obtained condition specific feedbacks from the MN to the GRN by identifying essential reactions in minimal media environmental conditions using FBA. We have used the augmented GRN so obtained, denoted $ \mathcal{G_C} $, to find modules in the combined GRN-MN.
This procedure can be automated once the following three inputs are available: (1) the GRN of an organism, (2) its MN along with an FBA model, and (3) a database of metabolite-TF interactions in the organism. These inputs are likely to be accessible for more and more organisms. Then the above method can be used to automatically find modules in the organism. The procedure does not find all modules but those that it does seem to have a fairly tight functional role in the organism.

\subsection{Module functionality and sign of constituent links}
Our method of finding modules by identifying the  strongly connected components of $ \mathcal{G_C} $ does not use information about the sign of the links (positive or negative) but only the directionality of the links. However in assigning functionality, the sign of the links is crucial. E.g., in Fig. \ref{fig:graphC_architecture_4node_modules}A the consumer motif functionality can be realized only via the two shown cases. If the sign of one link had been different, no simple functionality could have been inferred. For example, had the sign of the link from \textit{gutR} to \textit{gutP} been negative, the consumer motif functionality would not have been realized. Similarly, in other modules, the combination of signs of links from the source to the destination node is crucial for understanding module functionality. For a given directed subgraph, only a few combinations of the signs of links endow the subgraph with a simple (easy to identify) functionality. The fact that our construction, blind as it is to the sign of the links, finds subgraphs where the signs happen to be just right for assigning a simple functionality to the subgraph, is a non-trivial validation of the approach. It should be mentioned, however, that a module with a `wrong' sign might have a function that is more complex (e.g., a dynamical property that permits a better response to a time varying signal). Identifying  such functionalities requires a more detailed analysis that is beyond the scope of the present work. It is interesting that most modules have a sign combination that allows for an easy-to-interpret functionality; modules that are difficult to interpret functionally are rare (though there are a few, see SI \nameref{S5_ec_graphC_module_analysis}).

\subsection{Modules of \textit{E. coli} and \textit{B. subtilis}}
Each module represents a dynamical system in its own right. The circuit diagrams we have given represent a starting point for constructing Boolean or ordinary differential equation based models for these modules. We also provided evidence in section \ref{sec:chap4_structure_of_core} that the core or LSCC of \textit{E. coli} had its own internal modular structure.

\subsection{Limitations and future directions}
\subsubsection{Robustness of the condensed graph and its biological interpretation}
A question arises as to whether the SCCs and the condensed graph we have constructed after including metabolic feedbacks are robust to future network change as databases expand. We have shown that almost all the SCCs other than the core have an identifiable functionality in the organism. In view of the fairly clear biological role we have identified for them, it is unlikely that future development of the databases will cause these SCCs to disappear, because the role these SCCs can perform in contributing to the organisms' fitness has been identified. We do expect that many of the SCCs will be enhanced to include other nodes and links in the future.
However this assertion cannot be made at this point about the core which controls the organism globally, and about the hierarchy in the new condensed graph we have obtained. We do not as yet have a very compelling logic for why some of the parts that are present in the core should appear there or why the hierarchical structure of the condensed graph should be what it is. This is a topic to be investigated in the future.

\subsubsection{Extensions of the approach}
\textbf{(a)} In our determination of modules of \textit{E. coli} and \textit{B. subtilis}, we included the feedbacks from MN. In the process we considered only a subset of the MN. That subset was composed of the essential reactions of the metabolic network. Taking only the part of the metabolic model that is constituted by the essential reactions is somewhat restrictive. It may be useful to go beyond the essential reactions in order to get better modules. We have tried using a flux vector which is obtained as a solution of the FBA (and contains more than the essential reactions) to obtain condition specific graphs. This in fact yields additional modules and also adds additional nodes and links to some existing modules. However since flux vectors are not unique (while essential reactions are) a systematization of this approach is required for further extension. \textbf{(b)} Similarly, as discussed at the end of the penultimate para in section \ref{sec:chap4_size_3_modules} in the context of the D-xylose module the inclusion of next to nearest neighbors in the metabolic graph will allso lead to some enlarged modules. \textbf{(c)} Enzyme inhibition: The GRN of an organism include genes coding for enzymes that catalyze the metabolic reactions. In our hierarchical picture these genes come in the level-0 (initiated schematically in red in Fig. \ref{fig:schematic_link_interpretation_and_grn_feedback_from_mn}B). The metabolites of the MN can bind allosterically to enzymes (in addition to TFs) and alter their activity. The effect of this on the GRN would be comparatively local because the genes that code for enzymes belong to level-0 of in our hierarchical picture of the GRNs, hence their influence is restricted. An example of such feedbacks is present in the tryptophan system in \textit{E. coli} which has been well studied and mathematical models have been developed for it (see \cite{yanofsky_rna-based_2007,santillan_dynamic_2004} and references therein). 
Our work does not include such feedbacks. We expect that such feedbacks would enrich the modules already obtained by bringing in the local elements of the metabolic network and thus provide a more complete picture of the modules.

\subsubsection{Dynamics of the cell}
Our work here is concerned with the structure of bacterial GRNs. A natural extension to it is the understanding of the dynamics through mathematical modelling and simulation. A good number of approaches exist for modelling GRNs \cite{karlebach_modelling_2008}, including Boolean and ordinary differential equation based methods. In any mathematical modelling of a dynamical system one builds upon an understanding of its structure. An insight about dynamical modelling of GRNs is provided by our approach: first individually model the SCCs of the GRN which have feedbacks and are thus more complex, then integrate them in the causal flow of the hierarchy. It may be useful to pursue this `divide and rule' strategy for a large system such as the GRN.

\section{Methods}
\subsection{Hierarchical organization of nodes}\label{methods_hierarchical_organisation}
\begin{figure}[h!] %
	\centering
	\includegraphics[width=.85\textwidth]{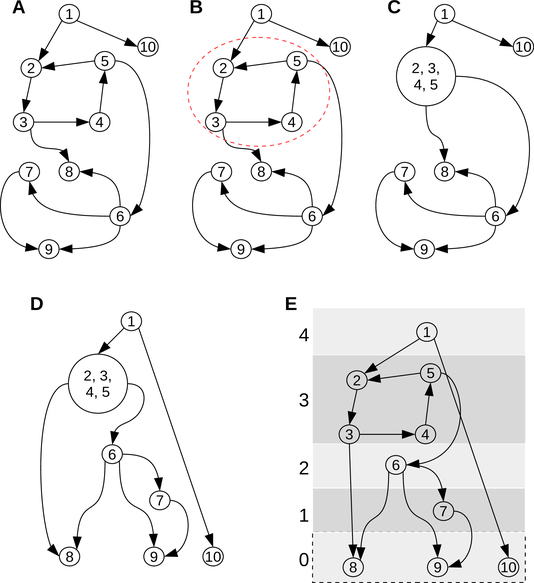}
	\caption{\textbf{Hierarchical decomposition} Demonstration of arranging the nodes of a toy network into hierarchical organisation used in this paper. The level 0 nodes have been enclosed in a dotted box to indicate that in our figures these nodes have not been shown due to their large number and keep focus on the details of regulation amongst the TFs.}
	\label{fig:methods_hierarchical_toy}
\end{figure}

We describe the procedure used in the paper to arrange the nodes of a network into hierarchical levels.
The algorithm is as follows: (1) Graph condensation. This involves identification of the SCCs of the graph, and substituting a single representative node for each SCC in place of its constituent nodes. This renders the graph acyclic. (2) Successive identification, assignment and removal of leaves (nodes with no out-degree). In the condensed graph, the leaves identified in the first iteration are identified as level 0 nodes. They are removed. The new leaves identified in the second iteration are identified as level 1 nodes. This process is repeated until all the nodes in the graph are assigned to a level.
We demonstrate this using a toy network. This procedure is a modification/combination based on works by Jothi et al \cite{jothi_genomic_2009} and Yu et al \cite{yu_genomic_2006}. Fig. \ref{fig:methods_hierarchical_toy} illustrates the procedure. 

Panel \textbf{A} of Fig. \ref{fig:methods_hierarchical_toy} shows the toy graph. It is a directed network consisting of 10 nodes and 12 edges. Panel \textbf{B} indicates the non-trivial SCC in the same network by enclosing nodes \circled{2}, \circled{3}, \circled{4} and \circled{5} of the network in a red dashed ellipse. Panel \textbf{C} shows the directed acyclic graph (DAG) obtained after the condensation of the toy network. The nodes forming the SCC  have been clubbed into one node represented by the bigger circle. Panel \textbf{D} shows the nodes of the condensed graph, i.e., the obtained DAG, arranged in a tree layout. The network in panel \textbf{D} can be used to assign nodes to different hierarchical levels. 

Panel \textbf{E} illustrates levels of the hierarchical  organization. The categorization of the nodes of the network into finer levels employs iterative leaf removal algorithm on the DAG in panel \textbf{D}. The leaves of a network are the nodes with no out-degree. The leaves are assigned to a level and then removed from the DAG which modifies the DAG and exposes new leaves. This procedure is iterated until all the nodes of the DAG are exhausted and the nodes assigned to subsequent levels. This is done as follows. In the DAG in panel \textbf{D}, the nodes  \circled{8}, \circled{9} and \circled{10} has no out-degree and are the leaves. We assign these nodes to level 0 (see panel \textbf{E}) and remove them along with their in coming links from the DAG. This leaves us with the modified DAG where \circled{7} is the new leaf. Then \circled{7} is assigned to the higher level, 1, and removed. This procedure is iterated assigning \circled{6} to level 2,  \circled{2,3,4,5} to level 3 and finally  \circled{1} to level 4.
The hierarchical levels along with the nodes are shown in panel \textbf{E} for the original toy network where we have expanded the SCC into its constituents nodes. In a GRN, level 0 generally contains many nodes and depending upon the need or for purposes of clarity, the level 0 might not have been shown in the hierarchical pictures used in the other sections of the paper.

\subsection{GRN with feedback from MN: $ \mathcal{G_B} $}\label{methods_grn_with_feedback_from_metabolism_gb}
We used the information about metabolic reactions inside the cell from publicly available genome scale metabolic models of \textit{E. coli} \cite{reed_expanded_2003} and \textit{B. subtilis} \cite{henry_ibsu1103:_2009}. The common genes between the GRN and corresponding metabolic model were used to place an edge from genes of the GRN to the metabolites of the reactions catalyzed by corresponding proteins. The metabolic model of \textit{E. coli}, iJR904 \cite{reed_expanded_2003}, contains 904 genes, 931 reactions and 761 metabolites, while the metabolic network of \textit{B. subtilis}, iBsu1103 \cite{henry_ibsu1103:_2009}, contains 1103 genes, 1437 reactions and 1381 metabolites. The influence of metabolites on TFs were obtained from RegulonDB \cite{salgado_regulondb_2001,salgado_regulondb_2013} and Ecocyc \cite{keseler_ecocyc:_2013} for \textit{E. coli}, while for \textit{B. subtilis} this information was obtained from Goelzer et al \cite{goelzer_reconstruction_2008}. 
We selected the reactions from the respective metabolic models (\textit{i}JR904 for \textit{E. coli} and \textit{i}Bsu1103 for \textit{B. subtilis}) which were catalyzed by the enzymes whose coding genes were also present in the GRN (i.e, graph $ \mathcal{G_A} $), and also whose metabolites altered the activity of the transcription factors.
In \textit{E. coli} we found 320 such genes catalyzing reactions involving 66 metabolites which altered the activity of 57 transcription factors. This introduced 462 links from its GRN ($ \mathcal{G_A} $) to the MN (from 320 genes to 66 metabolites) and 77 links from the MN back into the GRN (from 66 metabolites to 57 TFs).
In \textit{B. subtilis} we found 168 genes catalyzing reactions involving 29 metabolites which altered the activity of 24 TFs. This introduced 416 links from its GRN ($ \mathcal{G_A} $) to the MN (from 168 genes to 29 metabolites) and 34 links from the MN back into its GRN (i.e., from 29 metabolites to 24 TFs). 
We now constructed the augmented GRN graph $ \mathcal{G_B} $ from $ \mathcal{G_A} $ by adding new nodes corresponding to the TF binding metabolites (66 and 29 in the two organisms) and new directed links to and from these nodes. The incoming links to these nodes were from the enzyme coding genes whose gene product catalyzed the metabolic reactions of these metabolites (462 and 416 links). The outgoing links from these metabolites were to the genes coding for the TFs to which these metabolites bind (77 and 34 links). Thus $ \mathcal{G_B} $ contains 66 new nodes and 462+77 new links over $ \mathcal{G_A} $ in \textit{E. coli}, and 29 new nodes and 416+34 new links in \textit{B. subtilis}. The metabolite-TF interactions are given in SI \nameref{S2_graphB}.
While the metabolic network has substrates (or metabolites) for a number of reactions carried out in the cell for different purposes, we make use of not all but only a subset of the substrates capable of altering the activity of the TFs. Thus, we are using not the whole of the metabolic network but only a part of it (as described below) which directly feeds back into the GRN.

\subsection{Construction of graph $ \mathcal{G_C} $}\label{methods_gc_construction}
\subsubsection{Determining the ECs and the corresponding essential reactions}
First, we consider the utilization of metabolic pathways by the organism in multiple environmental conditions (ECs). Each EC considered is characterized by a `minimal' medium in which there is a unique and limited source of carbon together with specified but unlimited other nutrients like nitrogen, phosphorous, sulphur, etc. An EC is characterized as aerobic if external oxygen is available, otherwise anaerobic. We use the computational technique of flux balance analysis (FBA) \cite{varma_metabolic_1994,orth_what_2010} to determine, for each EC, the reactions of the metabolic network that are essential for the organism to be able to grow in that medium (for details see section \ref{methods_fba}). 
The FBA simulation in a specified medium gives the maximal steady state growth rate of the organism in the medium and a `flux vector' that gives the steady state velocity of every reaction in the MN. An EC or medium is said to be viable if the maximal growth rate in it is positive. The model \textit{i}JR904 for \textit{E. coli} has 131 possible carbon sources and \textit{i}Bsu1103 for \textit{B. subtilis} 211 carbon sources. For \textit{E. coli} out of a total of 262 ECs considered (131 aerobic + 131 ananaerobic) only 89 produced viable minimal media in aerobic conditions and 69 in anaerobic conditions. Thus we had a list of 158 viable minimal media for \textit{E. coli}.  For \textit{B. subtilis} out of a total of 212 EC (all 212 aerobic, no anaerobic)  we had only 118 viable media, all aerobic. The list of viable minimal media is given in SI \nameref{S3_fba}. We denote the number of viable minimal media so obtained by $ M $ ($ M = 158 $ for \textit{E.coli} and 118 for \textit{B. subtilis}).

\subsubsection{Condition specific augmented GRNs and the graph $ \mathcal{G_C} $}
We then determine the essential reactions of the MN for each of the above ECs. 
An essential reaction in a metabolic model in a given viable EC, as the name suggests, is one, which, if blocked, results in zero simulated growth (i.e., blocking the reaction changes the EC from being viable to unviable). The determination of essential reactions under a given EC is done in the following standard way. The reaction to be tested for essentiality is first constrained to have a zero flux in the metabolic model, and then the metabolic model is simulated for growth in the particular EC using FBA. If this results in a zero growth value, then the reaction tested is an essential reaction in that EC, else it is not. This procedure is repeated for each reaction in the metabolic model and its essentiality is determined for the given EC. For example, in case of \textit{E. coli} in aerobic glucose minimal growth condition, FBA deemed 218 reactions out of 904 to be essential. Thus for a given EC, in augmenting the GRN with feedbacks from MN, instead of using all 462 links from enzyme coding genes to metabolites we only used the subset corresponding to essential reactions in glucose which were 71 in number. The metabolites that were not participating in the essential reactions were also excluded.
We determined the set of essential reactions under each of the 158 ECs for \textit{E. coli} and 118 ECs for \textit{B. subtilis}.  The essential reactions for each EC are listed in SI \nameref{S3_fba}. We augment the GRN with feedbacks from MN using a method similar to that employed previously to arrive at graph $ \mathcal{G_B} $, with an additional restriction that the reaction used to link the GRN and the MN must also be an essential reaction under any of the ECs simulated for growth via the metabolic model of bacteria.
For each of the ECs we generated a version of GRN augmented with feedbacks from the part of the MN constituted by essential reactions in that EC. This gave us 158 versions of GRNs augmented with condition dependent feedbacks from MN for \textit{E. coli}, and 118 versions for \textit{B. subtilis}.
We designate these instances of GRN with condition dependent allosteric feedbacks from metabolic network as graph $ \mathcal{G_C}_i $, where the index \textit{i} indicates that the graph is for a given growth condition, labelled by \textit{i}; \textit{i} goes from 1 to $ M $. 
SI \nameref{S3_fba} and \nameref{S2_graphB} together contain all the information to construct each $ \mathcal{G_C}_i $.
We next define the graph $ \mathcal{G_C} $ of an organism (\textit{E. coli} or \textit{B. subtilis}) to be union of graph $ \mathcal{G_C}_i $'s of that organism i.e., $ \mathcal{G_C} = \bigcup\limits_{i} \mathcal{G_C}_i $. The graph $ \mathcal{G_C} $ includes all the nodes and links that are present in any of the $ \mathcal{G_C}_i $. Another  equivalent way of arriving at the same graph $ \mathcal{G_C} $ is to first find the set of essential reactions in the MN for each EC, take the union of the sets of essential reactions across all ECs, find its intersection with the reactions corresponding to the links between enzyme coding genes and metabolites in $ \mathcal{G_B} $, and then use the latter set of reactions to augment the GRN with feedbacks from the MN. While the second procedure is a simpler way to arrive at graph $ \mathcal{G_C} $, the first procedure presents an opportunity to study the condition specific GRNs augmented with feedbacks from MN individually as well. The number of nodes and edges of various types in graph $ \mathcal{G_C} $ are listed in Table \ref{table:grn_and_mn_numbers_graph_C}. 

The graph $ \mathcal{G_C} $ is a sub-network of graph $ \mathcal{G_B} $. All the nodes other than metabolite nodes in $ \mathcal{G_B} $  are included in $ \mathcal{G_C} $ and their mutual links present in $ \mathcal{G_B} $ are also included in $ \mathcal{G_C} $. However, a metabolite node in $ \mathcal{G_B} $ belongs to $ \mathcal{G_C} $ if and only if the metabolite is produced or consumed in a reaction that is essential in any of the $ M $ media considered. A link from an enzyme node to a metabolite node in $ \mathcal{G_B} $ is included in $ \mathcal{G_C} $ if and only if the corresponding reaction catalyzed by the enzyme is an essential reaction in any of the $ M $ media considered.

The advantage of using essential reactions of a metabolic model under an EC lies in their unambiguous determination. An alternative approach (which we have explored but not discussed in detail here) is to use all the reactions with non-zero flux in a flux-vector. It is well known that in a given medium FBA has multiple flux vectors as solutions with the same maximal growth rate and the set of non-zero flux reactions is different for different flux vectors. This gives rise to an ambiguity in the set of reactions to be included if one works with all reactions with non-zero flux. However in a given medium the set of essential reactions is unique for a FBA model.

\subsection{Details of the performed FBA}\label{methods_fba}
We used the COBRA toolbox in Matlab \cite{schellenberger_quantitative_2011} to perform FBA 
with metabolic models of \textit{E. coli} \cite{reed_expanded_2003} and \textit{B. subtilis} \cite{henry_ibsu1103:_2009}. We simulate an environmental condition (EC) by (a) setting the lower and upper bounds of the carbon food source molecule defining the EC to be -10 and 0, respectively, (b) setting the lower and upper bounds of other required nutrients (CO$_2$, Iron, Water, H$ ^+ $, Potassium, Na$ ^+ $, NH$_4 ^+$, Phosphate, Sulphate) for growth to be -1000 and 0, respectively, (c) setting lower and upper bounds of other carbon molecules to be 0 and 1000, respectively, (d) setting lower and upper bounds of Oxygen to be -1000 and 1000 (for aerobic medium) or 0 and 1000 for anaerobic medium, respectively, (e) setting lower and upper bounds of ATP maintenance reaction to be 0 and 1000, respectively. The COBRA function \texttt{singleRxnDeletion} along with a growth rate cut-off of 1e-6 was used to determine the set of essential reactions for a particular EC. The set of viable food sources, and essential reactions under each EC is given in SI \nameref{S3_fba}.

\section{Supporting Information}

S1-S6 are excel files containing all network and FBA data required to reproduce the results mentioned in the paper, as well as circuit diagrams and other details of the modules found in \textit{E. coli} and \textit{B. subtilis}. Files S1-S6, and higher resolution images of figures can be made available upon request from the authors.

\subsection*{S1} \label{S1_graphA}
{\bf $ \mathcal{G_A} $, GRN of \textit{E. coli} and \textit{B. subtilis} without feedback from  metabolism.}  The excel sheets list the genes, interactions, regulators, sigma-factors, TFs, ncRNAs, regulated genes, details of SCCs, and hierarchical levels of the GRNs.

\subsection*{S2} \label{S2_graphB}
{\bf $ \mathcal{G_B} $, GRN of \textit{E. coli} and \textit{B. subtilis} with feedback from metabolism.}  Lists the genes, metabolites, regulatory interactions, links from the GRN to respective MNs, feedbacks from MN into the respective GRNs, details of SCCs, and hierarchical levels of the GRNs.

\subsection*{S3} \label{S3_fba}
{\bf Details regarding FBA of MN of \textit{E. coli} and \textit{B. subtilis}.} Contains information about flux balance analysis of the MN of \textit{E. coli} and \textit{B. subtilis}: potential carbon food sources, viable carbon food sources, essential reactions of the organism for each viable minimal medium \textit{i}.

\subsection*{S4} \label{S4_graphC}
{\bf $ \mathcal{G_C} $, GRN of \textit{E. coli} and \textit{B. subtilis} with condition dependent feedback from metabolism.} Has sheets related to the GRN augmented with environmental condition dependent feedbacks from MN, $ \mathcal{G_C} $: genes, metabolites, regulatory interactions, details of SCCs, and hierarchical levels of the GRNs.

\subsection*{S5} \label{S5_ec_graphC_module_analysis}
{\bf Regulatory modules of \textit{E. coli}.} The functional modules of \textit{E. coli} classified  according to their size (number of nodes) accompanied with respective proximal regulatory logic circuit diagrams, the list of environmental conditions in which they are active, and a discussion of their possible functional role.

\subsection*{S6} \label{S6_bs_graphC_module_analysis}
{\bf Regulatory modules of \textit{B. subtilis}.} This file contains the same information as in \nameref{S5_ec_graphC_module_analysis}, but for \textit{B. subtilis} instead of \textit{E. coli}.

\section*{Acknowledgments}
We would like to thank Varun Giri, Parth Pratim Pandey, Srikanth Ravichandran, Areejit Samal and Pooja Sharma for various discussions, and the International Centre for Theoretical Sciences, Bangalore for hospitality where part of this work was done. SK acknowledges the University Grants Commission, India for a Junior Research Fellowship and Senior Research Fellowship. SM acknolwedges a Senior Research Fellowship from the Department of Biotechnology, Government of India. SJ acknowledges the Addie and Harold Broitman Membership of the Simons Center for Systems Biology, Institute of Advanced Study, Princeton, USA. He also acknowledges grants from the Department of Biotechnology and the Science and Engineering Research Board, Government of India, and R\&D grants from the University of Delhi.

\nolinenumbers

%
%
%
%
%


\end{document}